\documentclass[floatfix, a4paper, ,aps,prl,twocolumn,superscriptaddress, nobibnotes, full]{revtex4-1}

\usepackage{graphicx}
\usepackage[usenames,dvipsnames]{xcolor}
\usepackage{float}
\usepackage{mathtools}
\usepackage{amsmath}
\usepackage{amssymb}
\usepackage{natbib}
\usepackage{color}
\usepackage{psfrag}
\usepackage{wasysym}
\usepackage{float}
\usepackage{epstopdf}
\usepackage[normalem]{ulem}
\usepackage{textcomp}
\usepackage{hyperref}
\usepackage[utf8]{inputenc}
\usepackage[english]{babel}
\usepackage[notextcomp]{stix}
\usepackage{geometry}
\newcommand{\abb}{RCF~}
\newcommand{\monostable}{\textcolor[rgb]{1, 0.2, 0.2}{$\lgwhtcircle$}}
\newcommand{\axial}{\textcolor[rgb]{0.13, 0.55, 0.98}{$\vrectangleblack$}}
\DeclareRobustCommand\bent{\rotatebox{60}{\textcolor[rgb]{0.45, 0.98, 0.53}{$\vrectangleblack$}}}

\DeclarePairedDelimiter\abs{\lvert}{\rvert}%
\makeatletter
\let\oldabs\abs
\def\abs{\@ifstar{\oldabs}{\oldabs*}}

\hypersetup{pdfauthor={Bende et al.},pdftitle={How a bendy straw holds its shape}}

\begin{document}

\title{Overcurvature induced multistability of
linked conical frusta: How a `bendy straw' holds its shape}

\author{Nakul P. Bende}
\affiliation{Polymer Science and Engineering, University of Massachusetts Amherst, MA, USA}
\author{Tian Yu}
\affiliation{Department of Biomedical Engineering and Mechanics, Virginia Polytechnic Institute and State University, Blacksburg, VA, USA}
\author{Nicholas A. Corbin}
\affiliation{Department of Biomedical Engineering and Mechanics, Virginia Polytechnic Institute and State University, Blacksburg, VA, USA}
\author{Marcelo A. Dias}
\affiliation{Department of Engineering, Aarhus University, Aarhus, Denmark}
\author{Christian D. Santangelo}
\affiliation{Department of Physics, University of Massachusetts Amherst, MA, USA}
\author{James A. Hanna $^{ \ast}$}
\affiliation{Department of Biomedical Engineering and Mechanics, Virginia Polytechnic Institute and State University, Blacksburg, VA, USA}
\affiliation{Department of Physics and Center for Soft Matter and Biological Physics, Virginia Polytechnic Institute and State University, Blacksburg, VA, USA}
\author{Ryan C. Hayward $^{ \ast}$}
\email[To whom correspondence should be addressed: ]{hayward@umass.edu, hannaj@vt.edu} 
\affiliation{Polymer Science and Engineering, University of Massachusetts Amherst, MA, USA}

\date{\today}

\begin{abstract}
\noindent We study the origins of multiple mechanically stable states exhibited by an elastic shell comprising multiple conical frusta, a geometry common to reconfigurable corrugated structures such as `bendy straws'. This multistability is characterized by mechanical stability of axially extended and collapsed states, as well as a partially inverted `bent' state that exhibits stability in any azimuthal direction. To understand the origin of this behavior, we study how geometry and internal stress affect the stability of linked conical frusta. We find that tuning geometrical parameters such as the frustum heights and cone angles can provide axial bistability, whereas stability in the bent state requires a sufficient amount of internal pre-stress, resulting from a mismatch between the natural and geometric curvatures of the shell. We analyze the latter effect through curvature analysis during deformation using X-ray computed tomography (CT), and with a simple mechanical model that captures the qualitative behavior of these highly reconfigurable systems.
\end{abstract}


\maketitle

Elastic shells that can undergo snap-through transitions between different mechanically stable states provide a powerful means to design shape programmable structures with adaptable form and function \cite{hu2015buckling}. Perhaps the simplest example is a thin hemispherical shell, which can be turned `inside-out' to yield a locally stable configuration. Numerous studies have aimed to understand the complex geometric and mechanical relationships underlying such behavior, and to thereby enable the design of reconfigurable elastic structures \cite{calladine1989theory, pippard1990elastic, seffen2006metal, pandey2014, reid2017geometry}. The resulting reconfigurable shells have found applications such as switchable micro-lenses \cite{crosby2007lens}, microfluidic pumps \cite{holmes2014pump}, and actuators \cite{shankar2013contactless, bartlett2016elastic, zirbel2016bistable}. Complementing these engineered examples, many biological mechanisms have been uncovered that exploit snap-through transitions between mechanically stable states of slender elastic structures to achieve rapid motion \cite{forterre2005venus, hayashi2009mechanics, hummingbird2011, 2013bacteria}.

\setcounter{figure}{0}
\begin{figure}[h]
    \centering
    \includegraphics[width=8.2 cm]{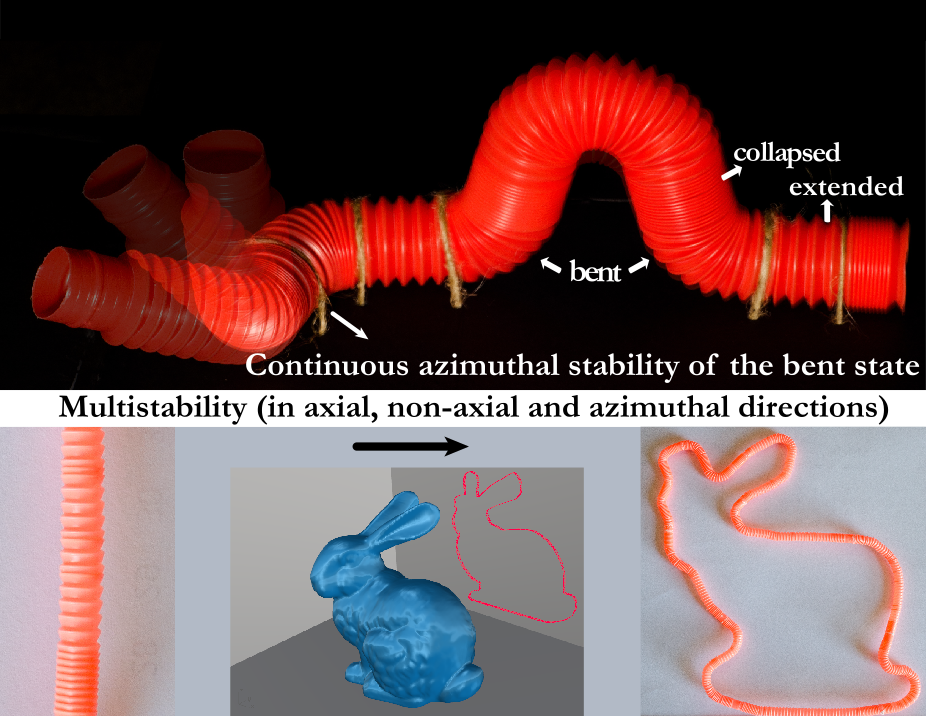}
    \caption{
    \textbf{(Top)} Multistability of a \textit{Pop Toob} demonstrated by deforming it into an arbitrary 3D space curve containing extended, collapsed, and bent subunits (overlaid). \textbf{(Bottom)} An initially straight section of bendy straw reconfigured into a planar projection of the Stanford bunny model \cite{stanford_bunny1994}.
    }
    \label{bendy_straw}
\end{figure}

While much of the literature has focused on {\it bi}stability, systems that support {\it multiple} stable states are attractive for the design of highly reconfigurable structures. Interestingly, such multistability can often be found in the simple motif of corrugated cylindrical shells exemplified by a `bendy straw'. This motif consists of repeating units of two non-identical conical frusta in opposing orientations, usually connected at a thinner `crease'-like region (Fig.\ref{bendy_straw}). Multistability in this specific geometry is manifested in the following two ways. Switching between extended and collapsed states through a full inversion of one frustum, referred to here as `axial bistability', changes the length of the structure. Additionally, such shells often show stability in a partially inverted (non-axially symmetric) `bent' state, where the bending direction can be continuously varied while preserving mechanical stability. The versatility of this architecture can be seen in Fig.\ref{bendy_straw}, where an arbitrary three-dimensional space curve and a planar projection of the Stanford bunny \cite{stanford_bunny1994} are formed. Similar geometries can be found in other commercial products such as transport ducts and collapsible camping containers, but the patent literature fails to explore the fundamental mechanisms underlying multistability \cite{friedman1951flexible, harp1968, mikol1989adjustable, 1975tubular, 2010collapsible}. 

\setcounter{figure}{1}
\begin{figure*}[t]
    \includegraphics{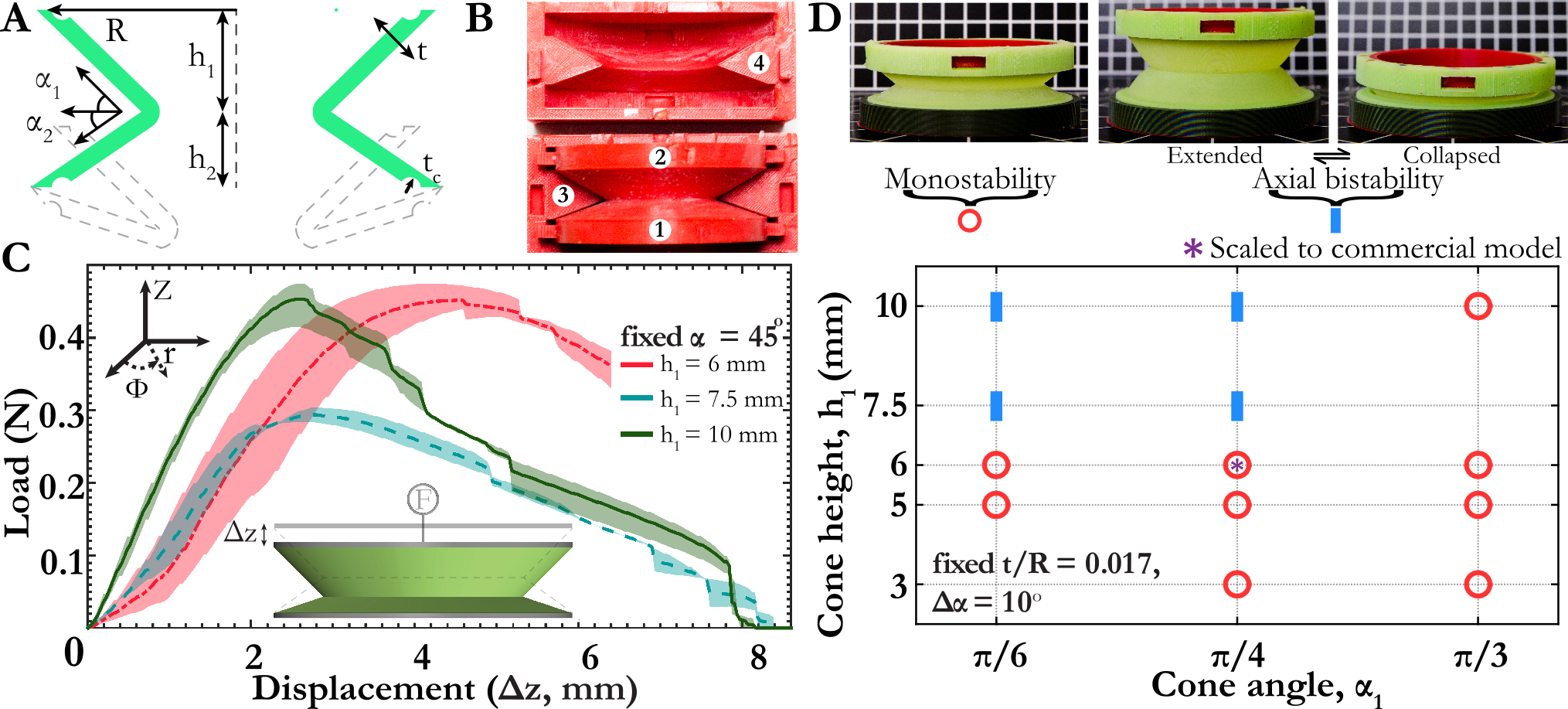}
    \caption{
    \textbf{(A)} \abb geometry (collapsed state shown in dashed grey).
    \textbf{(B)} Four-part, 3D-printed negative molds used for curing \abb using poly(vinyl siloxane) elastomer.
    \textbf{(C)} Response under axial load for samples with fixed $\alpha_1 \textrm{ = } 45^o \textrm{, } \alpha_2 \textrm{ = } 35^o$, and varying $h_1$. Shaded regions show the max/min values around the mean for two runs on three different samples each.
    \textbf{(D)} Stability of elastic \abb with varying $h_1$ and $\alpha_1$, recorded for at least 3 samples for each point shown, denoted as monostability (\monostable), and axial bistability (\axial).
    }
    \label{sample_prep}
\end{figure*}
Axial bistability between extended and collapsed states for each unit of the structure can be understood in much the same manner as for the hemispherical shell described above. Full inversion of a conical frustum leads to a nearly isometric shape devoid of stretching except at the boundary with the neighboring cone, whereas states intermediate between the two mirror symmetric shapes require stretching of the cone itself. Since the material in the crease is thinner, or otherwise weakened, compared to the rest of the shell, this means that the fully inverted state represents a local minimum in stretching energy, thereby providing mechanical stability as long as the bending energy cost for inversion is not too large (i.e., the shell is sufficiently thin compared to its overall dimensions) \cite{bende2015snap}. However, mechanical stability in the partially inverted state is not as easily understood, yet is vital to the utility of these highly reconfigurable structures.   

\section*{Geometry and fabrication}

To better understand the behavior of the bendy straw, we simplify the geometry to a single pair of non-identical opposed `reconfigurable conical frusta' (RCF) as shown in Fig.\ref{sample_prep}A. The two frusta are defined by the shell thickness ($t$) and base radius ($R$), as well as the cone heights ($h_1$ and $h_2$) and slant angles ($\alpha_1$ and $\alpha_2$) for both the `upper' and `lower' frustum. Note that one height or angle is constrained by the fact that $R$ is the same between the two frusta. We introduce creased regions (fixed at $t_c \textrm{ = }0.5t$) at the conical bases to guide inversion along this boundary and approximate the thinned portion found in commercial products. We note that the boundary constraints on the two conical bases are important for multistability, and are provided by the neighbouring units in a bendy straw. Here, we impart similar constraints to \abb by connecting the base of each cone to a thick (5 mm) cylindrical shell, which forces these boundaries to remain circular and facilitates gripping during mechanical characterization. We reduce the remaining five-dimensional parameter space for the \abb ($t \textrm{, } R \textrm{, } h_1 \textrm{, } \alpha_1 \textrm{, } \alpha_2$) by choosing several parameters to match a toy model (\textit{Pop Toob}, Poof-Slinky Inc., USA), i.e., $t/R \textrm{ = } 0.017$ (with absolute dimensions increased two-fold to $t = 0.5$ mm) and $\Delta\alpha \equiv \alpha_1 - \alpha_2 \textrm{ = }10^o$, leaving a two-dimensional parameter space spanned by $h_1$ and $\alpha_1$. 

To fabricate elastic \abb, we use a room temperature curable, two-component poly(vinyl siloxane) elastomer (PVS, Elite double 32, Zhermack Inc., Italy; $Y$ = 1.36 MPa) for its ease of handling and consistent material properties. We impart the proper \abb geometry to PVS shells by designing 3D printed, four-part negative molds (Fig.\ref{sample_prep}B), inside which the PVS mixture can be cured (after degassing) to obtain \abb with the desired geometries (\hyperlink{SI}{see SI}).

\section*{Characterizing \abb stability}

We establish the stability of fabricated elastomeric \abb by manipulating them into each shape (extended, collapsed, and bent) by hand, with further quantification provided by load/displacement measurements -- all carried out with the conical bases rigidly constrained to remain circular using 3D printed grips (see SI). We begin by testing stability for a scaled model of the \textit{Pop Toob} ($h_1 \textrm{ = } 6 \textrm{ mm, } \alpha_1 \textrm{ = } 45^o$), which surprisingly shows only monostability (denoted by \monostable, Fig.\ref{sample_prep}D). This is further reflected in the axial load-displacement curve (red curve, Fig.\ref{sample_prep}C), which initially shows a monotonic increase in force, followed by a softening when the lower frustum inverts, without the load ever dropping to zero. Exploring the parameter space further, we observe that setting $h_1 > 6$ mm and $\alpha_1 < 60^o$ results in elastic \abb with axial bistability (denoted by \axial). Here, the response of samples under axial displacement is characterized by a vanishing force around $\Delta z \textrm{ = } 8 \textrm{ mm}$, indicating a loss of contact between shell and indenter as the lower frustum snaps through to an inverted state (\hyperlink{SI_M}{Movie M1}). The appearance of axial bistability for a sufficiently large cone height is in accordance with the conventional understanding that a sufficiently thin shell is bistable between mirror inversions. However, all of the elastic \abb tested here lack stability in the bent state, failing to capture the multistability that makes the commercial products so useful for shape reconfiguration.
\setcounter{figure}{3}
\begin{figure*}[t]
    \centering
    \includegraphics[width=17.5 cm]{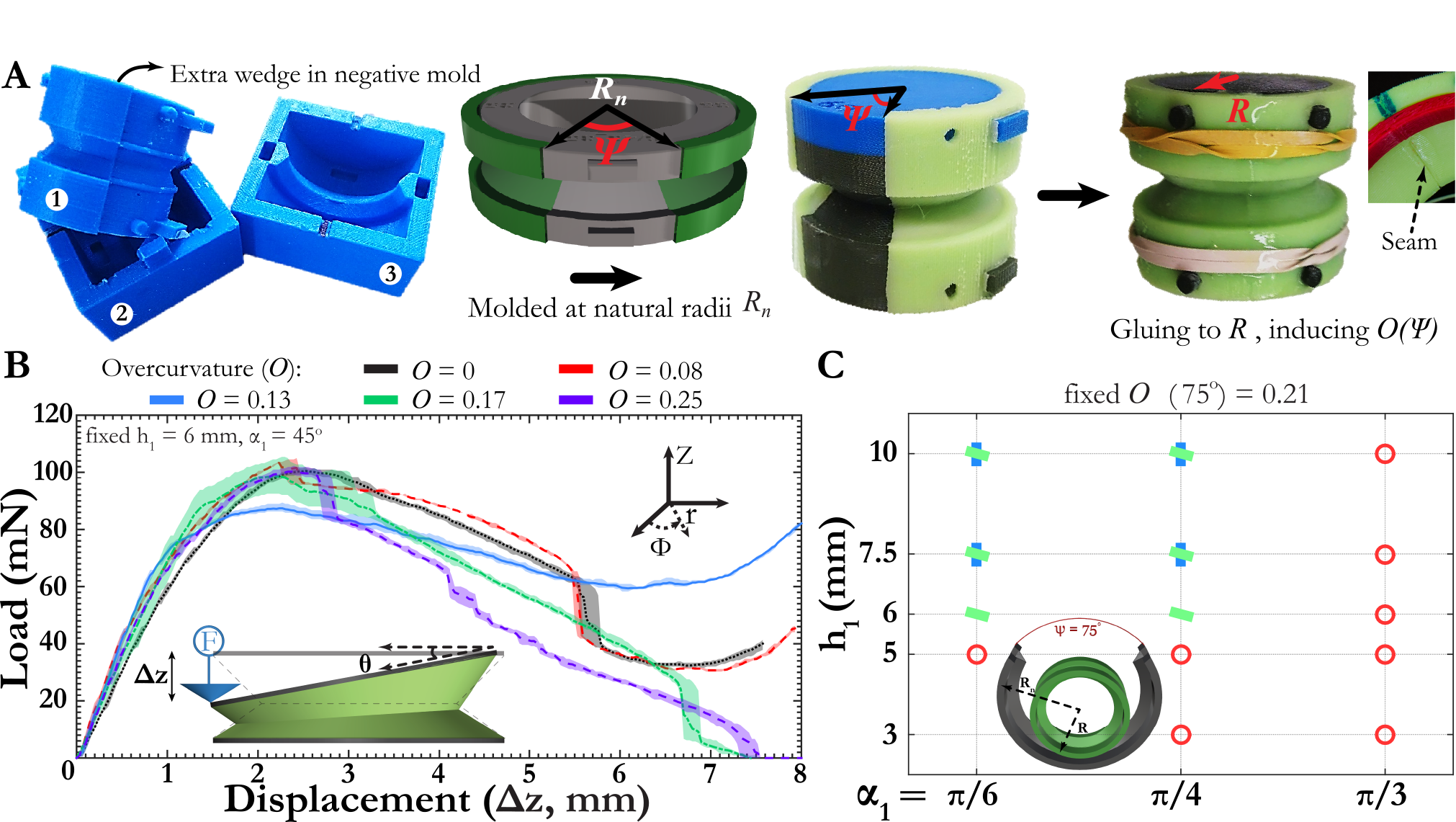}
    \caption{
    \textbf{Introducing overcurvature in \abb:}
    \textbf{(A)} Redesigned three-part molds for fabricating \abb with a wedge of arc angle $\psi$ missing.
    \textbf{(B)} The molded sample with natural radius $R_{n}$ is glued by curing uncrosslinked polymer to the intended radius $R$, introducing a controlled overcurvature $O(\psi)$.
    \textbf{(C)} Response of samples with $h_1 = 6$mm, $\alpha_1 = 45^o$, $\alpha_2 = 35^o$ and varying $O$ for a non-axial point load. Shaded regions show the max/min values around the mean for two runs each on two different samples.
    \textbf{(D)} A state diagram matching that in Fig. \ref{sample_prep}C, but now with pre-stress corresponding to $O = 0.21$, demonstrating that stability in the bent state  (\bent) is induced.
    }
    \label{builtin_stress}
\end{figure*}

\section*{Effect of geometrical frustration on \abb stability}

To further investigate the observed lack of stability in the bent state, we reconsider the commercial products that motivate our study. Remarkably, upon cutting axially along one side, the corrugated sections with original base radius $R$ relax by opening up to a natural base radius $R_n$ (Fig.\ref{stress}). 

This behavior, indicative of a built-in pre-stress, is consistent among several thermoplastic and elastomeric products, including \textit{Pop Toob}, a collapsible dog bowl (Roysili, USA), and bendy straws. Relieving this pre-stress deprives the structures of stability in both the inverted and bent states, leaving them with a smooth, accordion-like deformation, as shown in \hyperlink{SI_M}{Movie M2} for \textit{Pop Toob}. Gluing the structure back to the starting radius $R$ restores multistability. In contrast, closing the sample at its natural radius $R_n$ by gluing in an extra section from another unit restores the topology of the original sample without introducing pre-stress; such samples are also monostable. These observations suggest that sufficient pre-stress, as a result of curvature mismatch, is a necessary condition for stability in the partially inverted bent state, whereas our results on elastic RCF indicate that axial bistability can be achieved for samples of appropriate geometry without pre-stress.  

\setcounter{figure}{2}
\begin{figure}[H]
    \centering
    \includegraphics[width=8.5 cm]{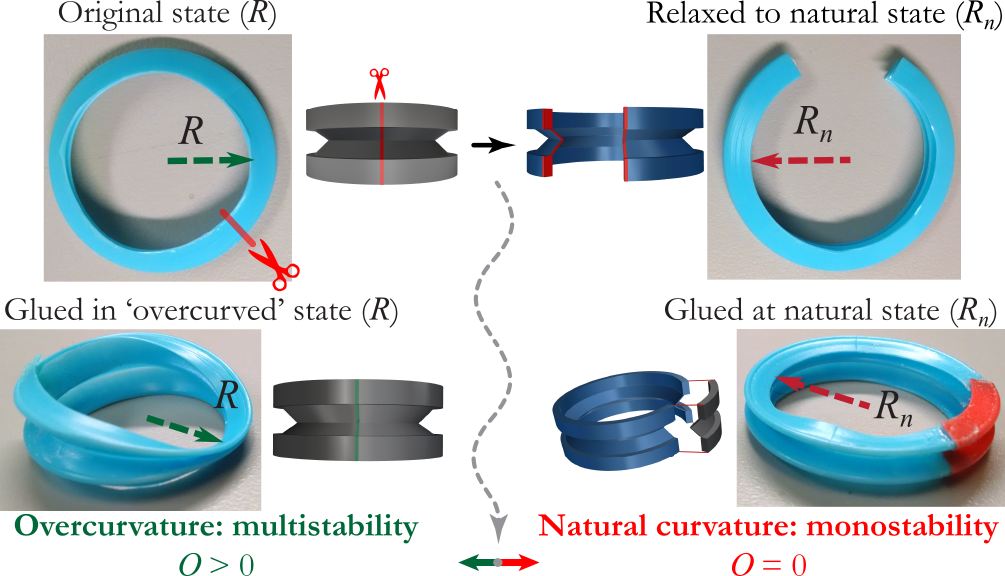}
    \caption{
    \textbf{Effect of overcurvature on multistability of a \textit{Pop Toob}:}
    Cutting a multistable sample opens the structure from an original curvature $R$ to a natural curvature $R_n$.  Gluing the unit back to its original overcurved state restores the multistability (lower left), whereas gluing the unit at the natural radius $R_n$ using an extra piece fails to do so (lower right).}
    \label{stress}
\end{figure}

Stress induced by geometrical frustration is known to affect the mechanical equilibria of complex rods \cite{mouthuy2012overcurvature, dias2015wunderlich, audoly2015buckling}, sheets \cite{Efrati09JMPS} and shells \cite{kebadze2004bistable, seffen2011prestressed, hamouche2016basic}. To explore similar effects of incompatible curvature on the stability of conical frusta, we fabricate elastic \abb with controlled overcurvature by redesigning the negative mold ($\#$ 1, Fig. \ref{builtin_stress}A) to include an extra wedge of angle $\psi$, thus yielding samples with a portion of shell missing. These incomplete shells are molded with natural base radii $R_{n}(\psi) \equiv R/ (1 - \psi/2\pi)$, and then glued closed to radius $R$ by bringing the free edges together and curing additional silicone elastomer applied to the seam. This introduces an internal pre-stress due to the curvature mismatch, or `overcurvature', quantified as
\begin{equation}
    O(\psi) \equiv 1 - \dfrac{R}{R_{n}(\psi)} \textrm{ = } \frac{\psi}{2\pi}.
    \label{overcurvature}
\end{equation}
\vspace{0.5em}
Introducing pre-stress in this manner slightly increases the dimensions in the axial direction (by $< 10\%$), but offers a simple mechanism for encoding overcurvature in samples with otherwise similar geometry ($R$, $t$, $t_c$, $\alpha$, $h$). To study this effect, we program \abb in the previously explored parameter space, but now with $O \textrm{ }\equiv \textrm{ } 0.08 \textrm{ (}\psi \textrm{ = } 30^o), \textrm{ } 0.13 \textrm{ (}45^o), \textrm{ } 0.17 \textrm{ (}60^o), \textrm{ } 0.21 \textrm{ (}75^o) \textrm{ and } 0.25 \textrm{ (}90^o)$, and measure their stability in the bent state. 

A reliable test of stability in the bent state requires a change in tilt angle $\theta$ during force-displacement measurements (Fig.\ref{builtin_stress}B). We allow for this tilt using linear displacement along the axial (z) direction of an indenter placed at the edge of the shell and allowed to slide freely in the radial (r) direction (\hyperlink{SI}{see SI} for CAD schematics and details). Using this non-axial indentation setup, we revisit the scaled \textit{Pop Toob} geometry ($h_1 \textrm{ = } 6 \textrm{ mm}$, $\alpha_1 \textrm{ = } 45^o$), but now with controlled amounts of overcurvature. As shown in Fig.\ref{builtin_stress}B, all samples initially show a similar force response: an initial linear regime followed by a peak in force at $\Delta z \approx 2 \textrm{ mm}$. Subsequent indentation reveals the effect of overcurvature. Samples with $O \textrm{ = } 0.17 \textrm{ and } 0.25$ clearly undergo a snap-through transition to a stable bent state, whereas those with lower overcurvature ($O \textrm{ = } 0.08 \textrm{, } 0.13$) follow a similar force response to the control sample lacking  overcurvature (see \hyperlink{SI_M}{Movie M3}). Following this successful attempt at introducing stability in the bent state for one geometry via pre-stress, we re-investigate the parameter space in Fig.\ref{sample_prep}C. We fabricate each geometry in an overcurved state corresponding to $O = 0.21$, and find that all samples with $h_1 \geq 6 \textrm { mm and } \alpha_1 \textrm{< } 60^o$ exhibit stability in the bent state upon introduction of overcurvature (denoted by \bent, Fig. \ref{builtin_stress}C). Samples with lower values of $h_1 \textrm{, or } \alpha_1 = 60^o$, show no change in stability even at the highest level of overcurvature tested ($O \textrm{ = } 0.25$). The partially inverted bent state, whenever stable, is azimuthally degenerate and can be reconfigured in any direction, with the exception of positions near the glued seam where symmetry is broken. Interestingly, samples that initially lacked axial bistability did not gain it through overcurvature, again highlighting that the fundamental requirements for stability in inverted and bent states may be different. 

\section*{A simple model to capture the effects of geometrical frustration on stability}

The creation of new metastable energy minima through frustration is a well known concept, as embodied in canonical examples such as the Ising model on a triangular lattice \cite{isling_model_2006}. To understand how a geometric incompatibility such as overcurvature can lead to multistability, we consider a toy model that captures key qualitative features of \abb. This takes the form of a planar four-bar linkage with two torsional springs (Fig.\ref{4model}A) between a rigid ground link (length $2R$, dashed) and two rigid crank links (length $W$, shades of red) that prefer rest angles $\tau_l = \alpha_2$ and $\tau_r = (\pi-\alpha_2)$, on the left and right, respectively. Incompatibility is introduced through the disagreement between these rest angles and the length of the rigid floating link ($R - W \cos \beta$, in blue). This seeks to impose a `cone' angle $\beta$ different than that of the torsional springs ($\alpha_2$). Our parameter space is spanned by this mismatch $\beta-\alpha_2$, the original `cone' angle $\alpha_2$, and the width $W$ encoded in the length of the crank links. The possible stable states for this model are the extended state (e) and those obtained from it by full inversion about the ground link to a collapsed (c) state or partial inversion to yield a pair of bent states (e). Four-bar linkages with a single torsional spring were examined by Jensen and Howell, who found conditions for bistability of such structures \cite{jensen2003identification}. A second torsional spring allows for the introduction of frustration or pre-stress, and has also been studied for bistable \cite{jensen1999design} and tristable \cite{pendleton2008compliant} four-bar linkages, as pseudo-rigid models of compliant mechanisms. 

\setcounter{figure}{4}
\begin{figure}[h]
	\centering
	\includegraphics[width=8.5 cm]{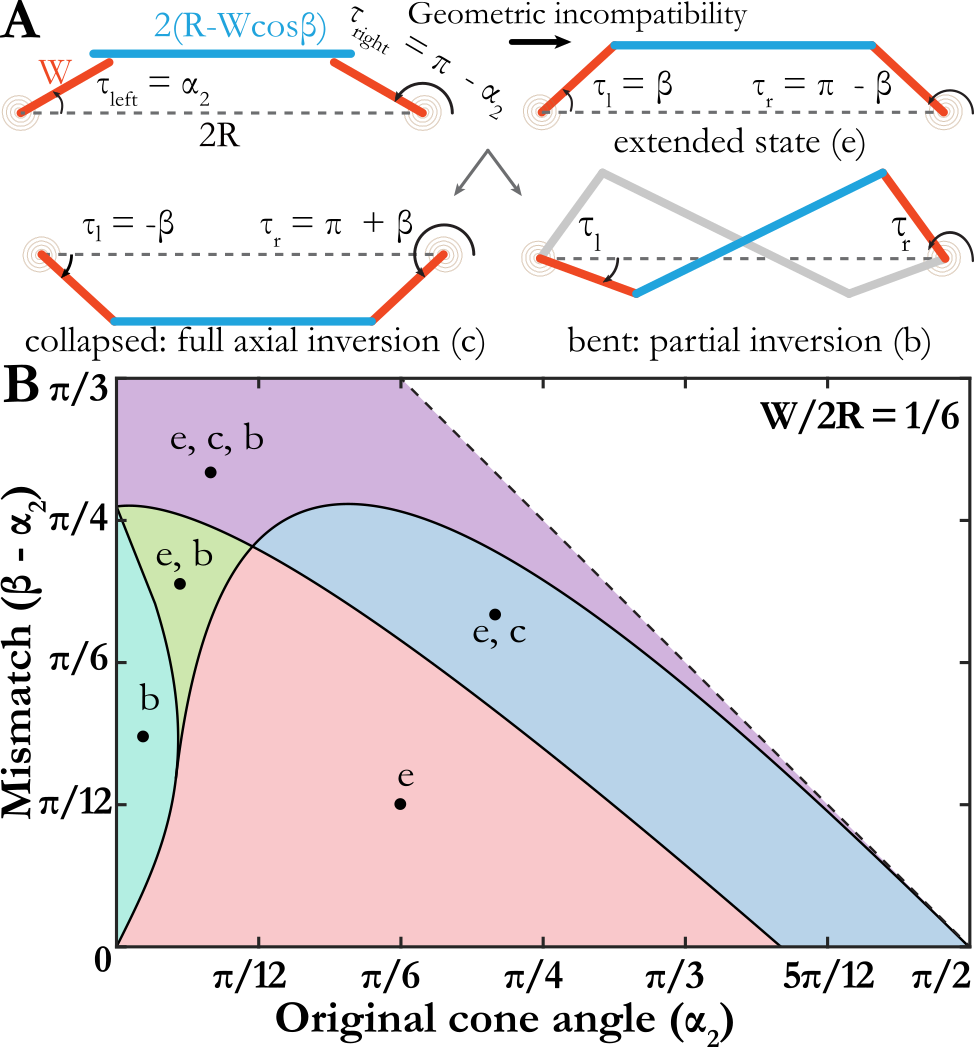}
	\caption{
	\textbf{Incompatible four-bar system:} \textbf{(A)} Construction, possible extended, collapsed, and bent states of an incompatible rigid four-bar linkage and torsional spring model.  \textbf{(B)} Stability of the model for a given link geometry as a function of rest angle and incompatibility.
	}
	\label{4model}
\end{figure}

For our model, we can write the Lagrangian as (complete derivation in \hyperlink{SI}{SI}),
\begin{align}
  \mathcal{L}=&\frac{1}{2} K_r (\tau_l - \alpha_2) ^2 + \frac{1}{2} K_r (\pi - \tau_r - \alpha_2) ^2 +\nonumber\\
  &\lambda [(2R+W\cos \tau_r-W \cos \tau_l)^2+(W \sin \tau_r - W \sin \tau_l)^2\nonumber\\
  &-(2R-2W \cos \beta)^2],
  \label{Lagrangian}
\end{align}
where $K_r$ is a spring constant and $\lambda$ a multiplier keeping the crank links at the ends of the floating link. We examine this Lagrangian for incompatibilities $\beta - \alpha_2 > 0$ (analogous to overcurvature) and rest angles $0 \le \alpha_2 \le \pi/2$, as shown in Fig.\ref{4model}B. We observe one, two, three, or four stable configurations corresponding to the extended, collapsed, and the pair of bent states. 

While the overall behavior is not simple, the general trends are that an increase in geometric incompatibility stabilizes additional states, and that some incompatibility is required to achieve stability in the bent state (b), while stability in the extended (e) or collapsed (c) states can be achieved at zero mismatch. The choice of width $W$ shifts the stability regions, but does not have much of a qualitative effect. The manner in which new states and barriers emerge in the energy landscape is shown in the \hyperlink{SI}{SI}. Incompatibility induced stabilization of bent states can be understood by considering a linkage with rest angles $\alpha_2 \textrm{ approaching } 0$, corresponding to a nearly flat cone.  Incompatibility in the form of a large floating bar requires deviation from flatness, which is penalized quadratically in the spring energies, approximately $\tau_l^2 + \left(\pi - \tau_r\right)^2$ for a nearly-flat cone. Since the magnitudes $\abs{\tau_l}$ and $\abs{\pi - \tau_r}$ are both smaller in the bent state than in either the extended or collapsed states, the bent state becomes the ground state.  This region of mechanical stability of the bent state should persist in distorted form as the rest angles increase from zero.

\section*{In-situ analysis of deformation pathway for elastic \abb}

\begin{figure}[h]
    \centering
    \includegraphics[width=8.5 cm]{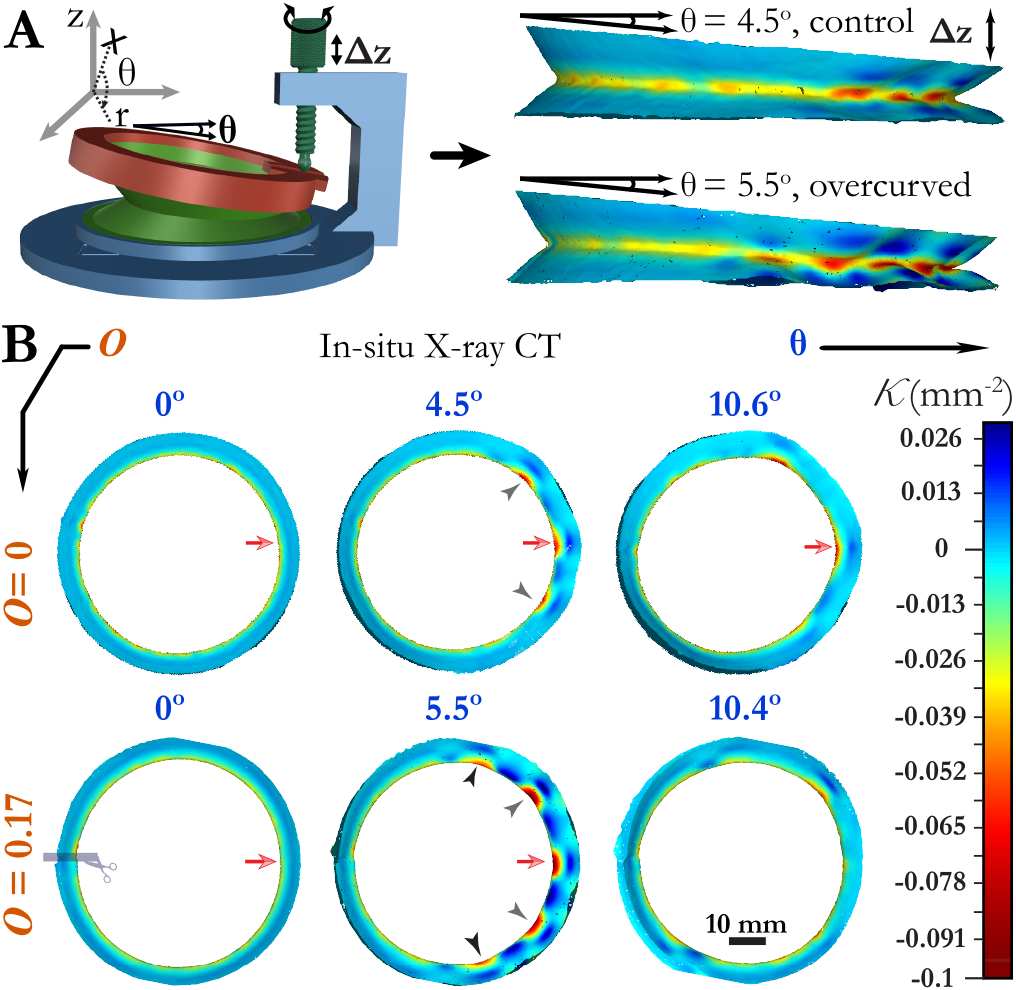}
    \caption{
    \textbf{Non-axial deformation of \abb in natural overcurved states:}
    \textbf{(A)} Schematics of the 3D printed harness for in situ X-ray CT studies. Side views of control ($O \textrm{ = } 0$) and overcurved ($O \textrm{ = } 0.17$) \abb shells at similar levels of tilt angle $\theta$ are also shown, with $\mathcal{K}$ projected.
    \textbf{(B)} $\mathcal{K}$ during non-axial loading for control [top row] and overcurved [bottom row] shells, as viewed from the top. Both shells possess similar geometrical parameters ($h_1 \textrm{ = } 6 \textrm{ mm, } \alpha_1 \textrm{ = } 45^o$), yet a higher magnitude of $\mathcal{K}$ is seen for the overcurved sample in the intermediate state.
    }
    \label{curvature_simulations}
\end{figure}

We return to the 3D shell structures to experimentally characterize their energy landscape by capturing the 3D profile of elastic \abb in situ during deformation using X-ray CT (IVIS SpectrumCT, Perkin Elmer, USA). Tomograms are recorded quasi-statically by fixing the sample in a deformed state using a custom harness, capable of imparting non-axially symmetric deformation in steps of $\Delta z \approx 0.1-1 \textrm{ mm}$ (see Fig.\ref{curvature_simulations}A). Curvature analysis on these 3D reconstructions provides Gaussian ($\mathcal{K}$) and mean ($H$) curvatures in the shells, details of which can be found in the \hyperlink{SI}{SI}. For thin conical shells, the presence of Gaussian curvature can be taken as a proxy for stretching, a costlier deformation pathway compared to bending.

To compare non-axial deformation with and without overcurvature, we plot $\mathcal{K}$ projected onto the surface of \abb ($h_1 \textrm{ = } 6 \textrm{ mm, } \alpha_1 \textrm{ = } 45^o$) fabricated with $O \textrm{ = } 0$ and $O \textrm{ = } 0.17$. In Fig.\ref{curvature_simulations}B, we compare tomograms in undeformed ($\theta \textrm{ = } 0^o$), intermediate ($\theta \approx 5^o$), and bent ($\theta \approx 10^o$) states. As expected in the undeformed state, we observe $\mathcal{K} \textrm{ = } 0$ in the conical regions and $\mathcal{K} < 0$ in the highly curved region where the cones attach to each other (see Fig.\ref{sample_prep}A). Upon indentation to $\theta \approx 5^o$, buckled regions with a higher concentration of $\mathcal{K}$ (`crests' marked with black and gray arrows) are seen near the point of indentation (red arrow). Throughout the deformation pathway, overcurvature leads to an increase in both the number of discernible crests and an increase in the magnitude of $\mathcal{K}$ associated with each crest (see side views). Indenting the sample further to $\theta \approx 10^o$ brings both of the \abb to the partially inverted configuration, though only the overcurved sample is stable in this bent state. The higher number of crests in the overcurved sample remains consistent throughout the entire non-axial indentation, supporting the idea that overcurvature increases the energetic cost of intermediate states enough to create an energy barrier that allows for stability in the partially inverted bent state.

\section*{Additive manufacturing of plastic \abb}

\begin{figure}[h]
    \centering
    \includegraphics[width=8.5 cm]{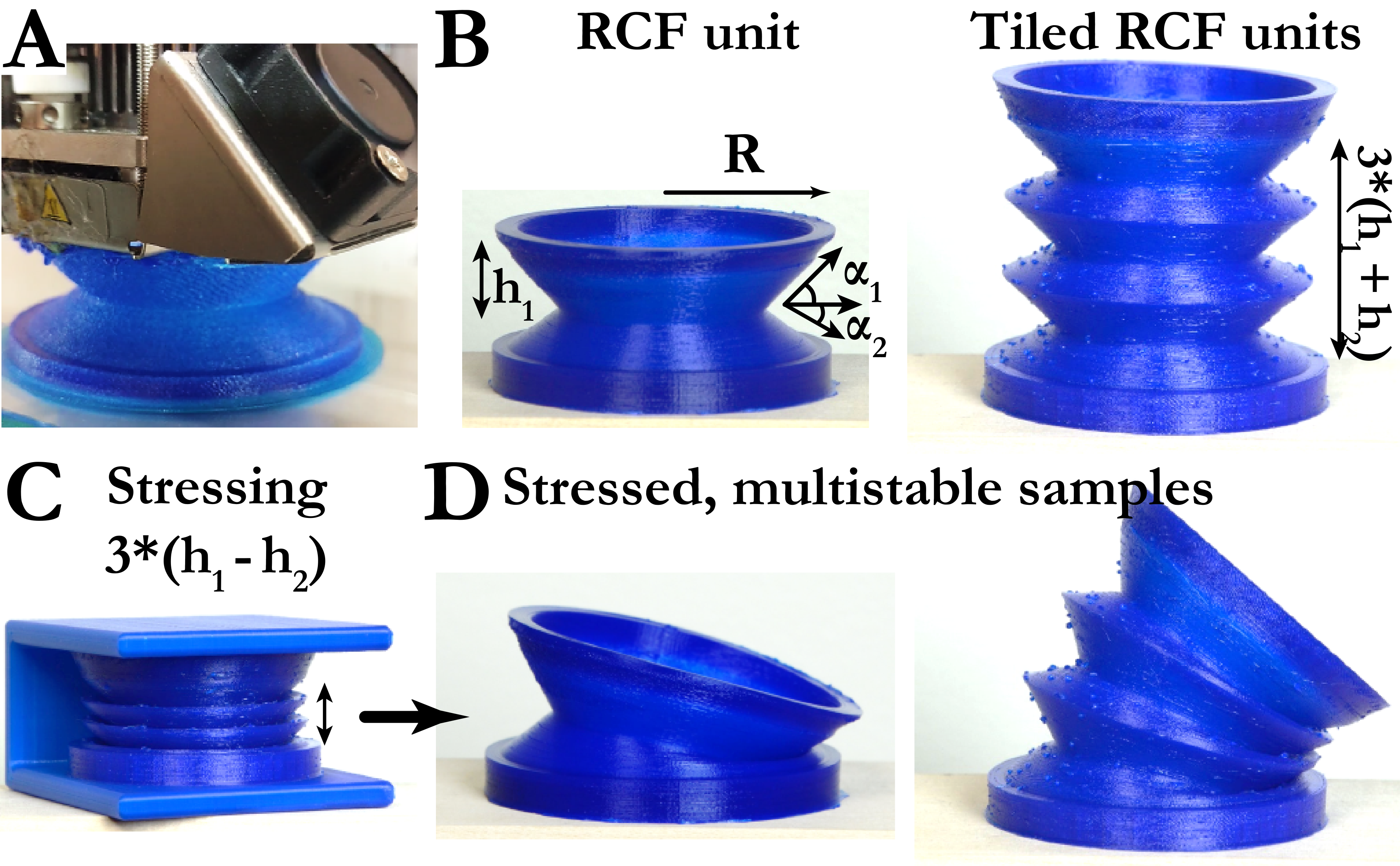}
    \caption{
    \textbf{3D printed \abb:}
    \textbf{(A)} 3D printing of an \abb using a poly(urethane) based elastomer.
    \textbf{(B)} Printed single-\abb and tri-\abb.
    \textbf{(C)} Fixing the elastomer shell under a collapsed state for $\approx$ 3 hours introduces a built-in residual stress for NinjaTek Cheetah.
    \textbf{(D)} \abb with built-in stress demonstrating stability in the bent state.
    }
    \label{3D_print}
\end{figure}

Finally, we present a simple protocol for fabrication of overcurved \abb out of viscoelastic polymers. Notably, the residual stress in an overcurved \textit{Pop Toob} is lost when it is left in an extended state for more than $\approx$ 1-2 days. Conversely, one can imagine harnessing viscoelastic behavior to induce overcurvature. As a proof-of-concept, we directly 3D print tiled \abb using a commercially available poly(urethane) based filament (Cheetah, NinjaTek, USA, using an Ultimaker 2 Extended+, Ultimaker, the Netherlands), as shown in Fig.\ref{3D_print}.

The as-printed \abb lack any overcurvature and are monostable. A controlled amount of overcurvature, and resulting multistability, in these samples can be introduced by constraining them into a collapsed state for a fixed duration (Fig.\ref{3D_print}C, \hyperlink{SI_M}{Movie M4}). The critical duration needed to induce sufficient overcurvature for multistability can be empirically found for specific materials by measuring the opening angle after cutting the sample open along the axis (\hyperlink{SI}{SI}, Fig. S5).  Leaving a sample in an uncompressed state for extended times is sufficient to reverse the process of imparting overcurvature. This result suggests that the use of viscoelastic relaxation to impart desired pre-stresses in shells may lead to more direct and rapid fabrication of more complex multistable structures from a wider range of materials.

\vspace{-1em}
\section*{Conclusions}

Structures with a high degree of geometric reconfigurability have recently garnered interest for use in surgical tools \cite{2013stiffflop}, deployable mechanisms \cite{filipov2016origami} and robotic arms \cite{kamrava2018programmable}. As a modular structure, the bendy straw offers an attractive architecture in such contexts, although the mechanism by which it supports multiple mechanically stable states has remained unknown until now. In this report, we explored the multistability of linked conical shells using experiments and a mechanical model, and established protocols for fabricating reconfigurable conical frusta with controlled overcurvature (pre-stress) in elastic and viscoelastic materials. We have established that axial stability between extended and collapsed states can be achieved through careful selection of geometrical parameters (cone height $h_1$, slant angle $\alpha_1$ and thickness $t$), while stability in bent, partially inverted, states requires a combination of appropriate geometry and sufficient pre-stress. The effect of overcurvature is to increase the stretching energy in the intermediate state during non-axial deformation, thereby providing an energy barrier and rendering the bent state metastable.

\section{Acknowledgements}
This work was funded by the National Science Foundation through EFRI ODISSEI-1240441. The authors thank Amy Burnside and Billye Davis for assistance with X-ray CT measurements, and Ishaan Prasad for useful discussions regarding the analysis of tomograms. JAH thanks D Vella for early discussions, and the hospitality of the Oxford Centre for Collaborative Applied Mathematics.

\newpage

\renewcommand{\thefigure}{S\arabic{figure}}
\renewcommand{\theequation}{S\arabic{equation}}
\setcounter{figure}{0}
\setcounter{equation}{0}

\hypertarget{SI}{\section*{{\huge Supporting Information:}}}

\section*{Protocol for designing and fabricating elastic \abb: }

\begin{figure}[h]
    \centering
    \includegraphics[width=0.35\textwidth]{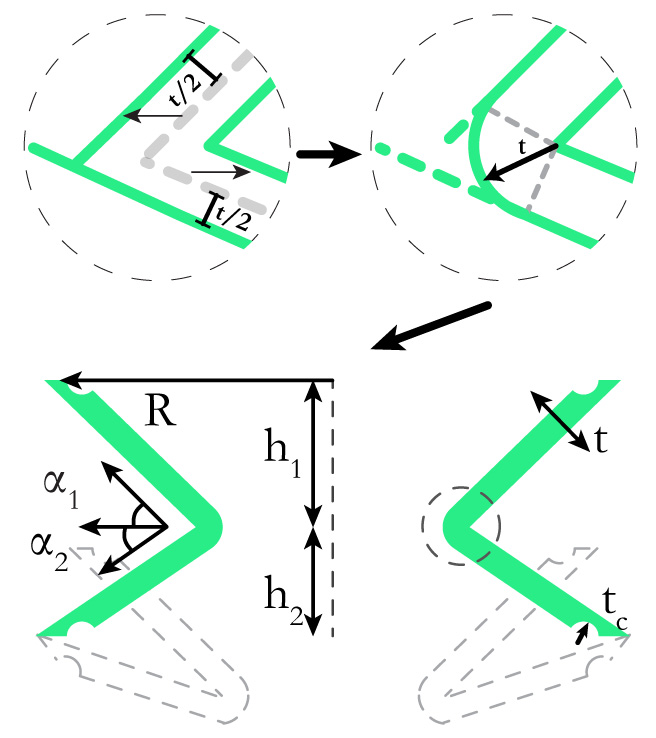}
    \caption
    {
    \textbf{\abb geometry: } Series of geometric operations necessary to obtain an \abb with given parameters.
    }
    \label{geometry}
\end{figure}

The re-configurable conical frusta (RCF) consist of two conical frusta inverted with respect to each other, as shown in Fig.\ref{geometry}. A conical frustum with base radius $R$, height $h_1$ and slant angle $\alpha_1$ (corresponding to an aperture angle of $\pi - 2\alpha_1$) constitutes the `upper' frustum. Fixing the slant angle of the `lower' frustum as  $\alpha_2 = \alpha_1 + \Delta\alpha$ and the base radius $R$ sets the cone height to $h_2 = R/ \tan(\alpha_2)$. The set of curves defining the mid-plane are then shifted laterally by $\pm t/2$ on either side to realize a shell with wall thickness $t$. 

Additional geometric operations necessary for fixing the shell at constant thickness $t$ are performed (Fig.\ref{geometry}), by shifting the bottom-inner walls inwards \{by $\tan(\alpha_1)-\tan(\alpha_2)$\} and curving the inner portion of both frusta around the central region. Lastly, the shell walls are `creased' on either end, in the form of a region around the bases with reduced thickness $t_c$.

\begin{figure}[h]
    \centering
    \vspace{0.8em}
    \includegraphics[width=0.5\textwidth]{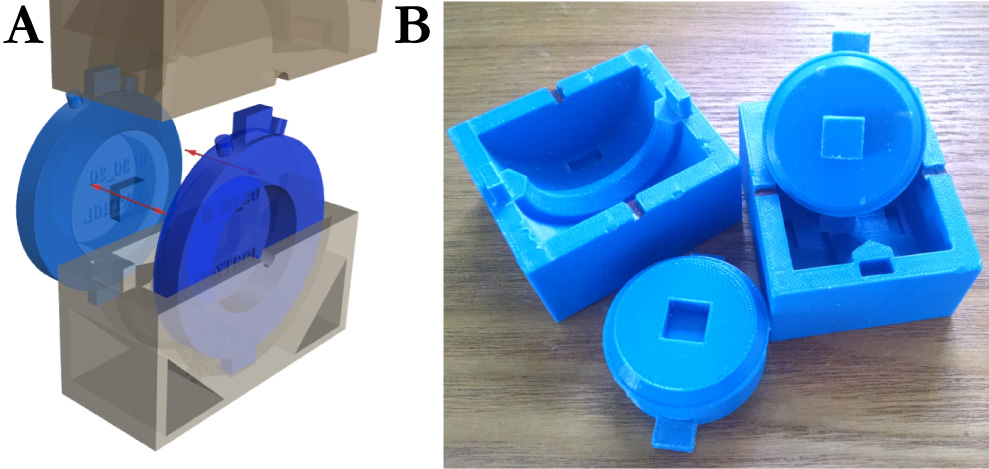}
    \caption
    {
    \textbf{(A)} CAD model, with indicated assembly for 4-part molds.
    \textbf{(B)} 3D printed negative molds used for fabricating \abb with a room temperature resin-cure molding of poly(vinyl siloxane) elastomer.
    }
    \label{CAD_molds}
\end{figure}

\noindent Elastic shells with intended geometries are fabricated using a two-part curable silicone elastomer, poly(vinyl siloxane) (PVS, Elite Double 32, Zhermack Inc., Italy), by resin cure molding at room temperature. The 4-part negative molds with appropriate geometrical parameters are modeled using a parametric 3D CAD algorithm (Grasshopper plug-in, Rhino, Robert McNeel Inc., USA) (Fig.\ref{CAD_molds}A). The molds are printed with poly(acrylonitrile-butadiene-styrene) using a commercial 3D printer (Dimension uPrint SE Plus, Stratasys, USA) as shown in Fig.\ref{CAD_molds}B.

\begin{figure}[h]
    \centering
    \vspace{0.8em}
    \includegraphics[width=0.5\textwidth]{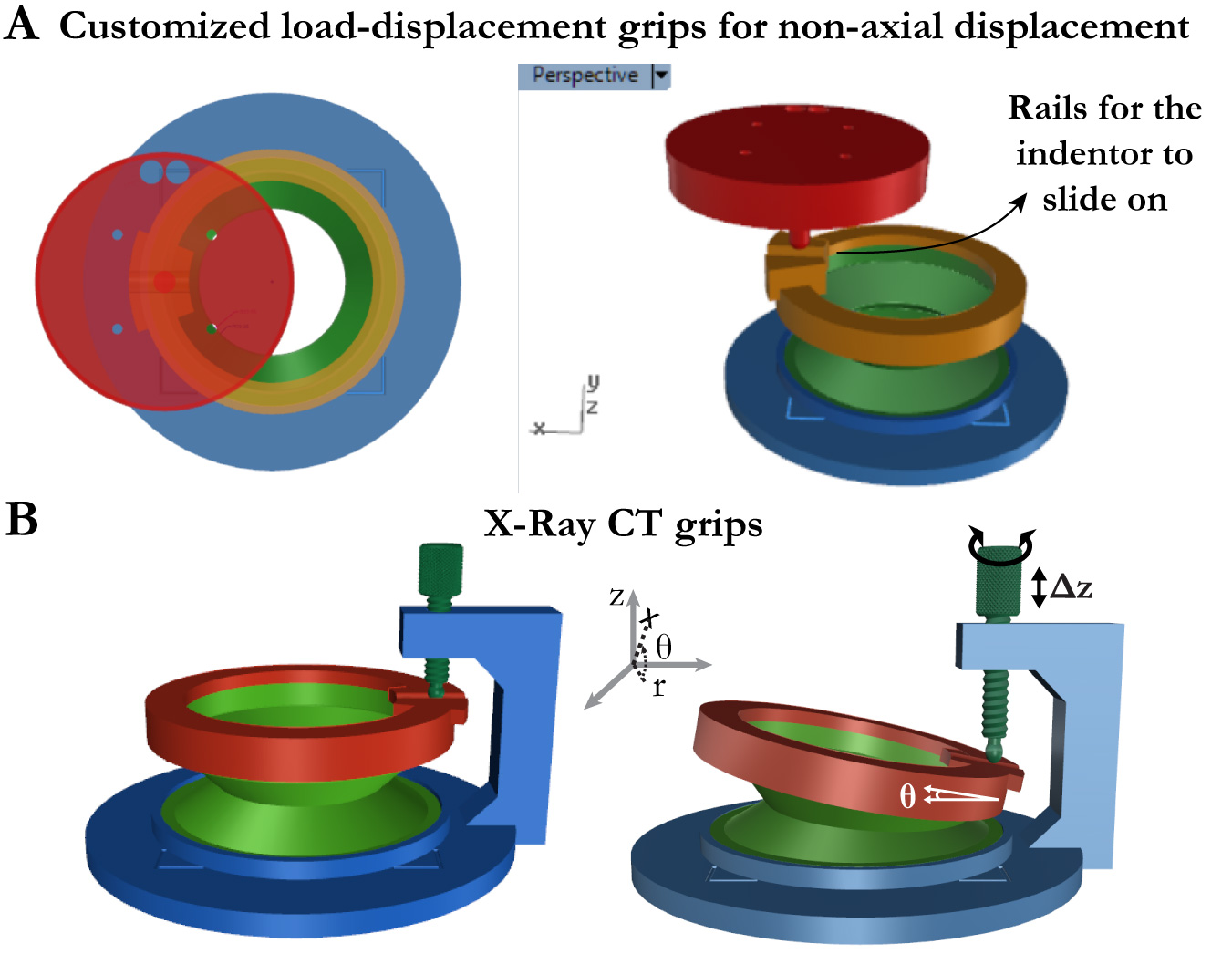}
    \caption
    {Customized indenter-grip combination for a controlled non-axial deformation of \abb for load-displacement setup \textbf{(A)} and X-ray CT setup \textbf{(B)}. 3D CAD models indicate the railings along which the indenter can slide freely, converting a linear translation in the $z$ direction to a controlled tilt angle $\theta$.
    }
    \label{non_axial}
\end{figure}
\begin{figure*}[t]
    \centering
    \vspace{0.8em}
    \includegraphics[width=0.8\textwidth]{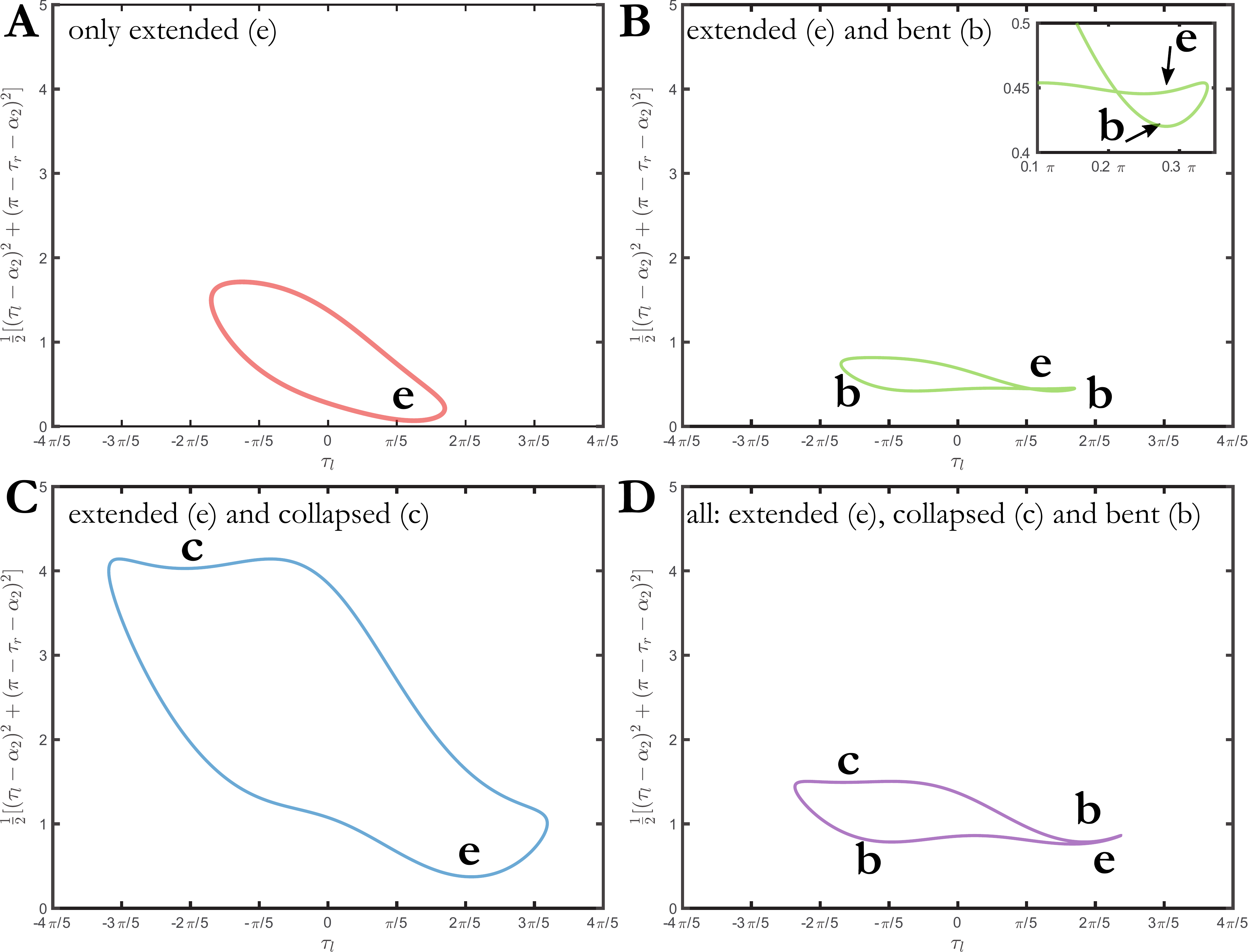}
    \caption
    {Energy landscapes for incompatible 4-bar linkage and torsional spring model. The curves correspond to samples with different geometrical parameters (cone angle: $\alpha_2$ and mismatch: $\beta-\alpha_2$) described in Fig.5 (main text).
    }
    \label{energy_4bar}
\end{figure*}
To cast an \abb geometry, a mold release agent (MR311, Sprayon, Canada) is first applied to 3D printed molds. A 1:1 (prepolymer:crosslinker) mixture of the PVS ingredients is introduced into the mold, followed by immediate degassing under vacuum for 8-10 min to remove any air bubbles. The molds are assembled as shown in Fig.\ref{CAD_molds}A, and held under room temperature for 30 minutes while curing. Finally, the molds are opened carefully and excess elastomer is removed from the final cured \abb. Samples with bubbles, irregular thickness or other defects are discarded. This method proves to be fairly robust, resulting in shells with $t = 0.5 \pm 0.05$ mm.

\section*{Custom indenter for non-axial deformation}
Imparting a controlled non-axial deformation to the \abb geometry poses a considerable challenge, as the point of contact travels in a curvilinear path. Limited by the ability to control only the linear translation of the indenter, a custom indenter-grip is designed for converting this linear translation in $z$ direction to non-axial tilt. As shown in Fig.\ref{non_axial}, a linearly translated indenter pushes on customized grips that hold the sample. At the point of contact, these grips consist of a smooth railing, allowing the indenter to slide in the lateral direction ($r$), converting the linear translation along the $z$ axis of the point indenter into a controlled non-axial tilt angle $\theta$.

\begin{figure*}[t]
    \centering
    \includegraphics[width=0.8\textwidth]{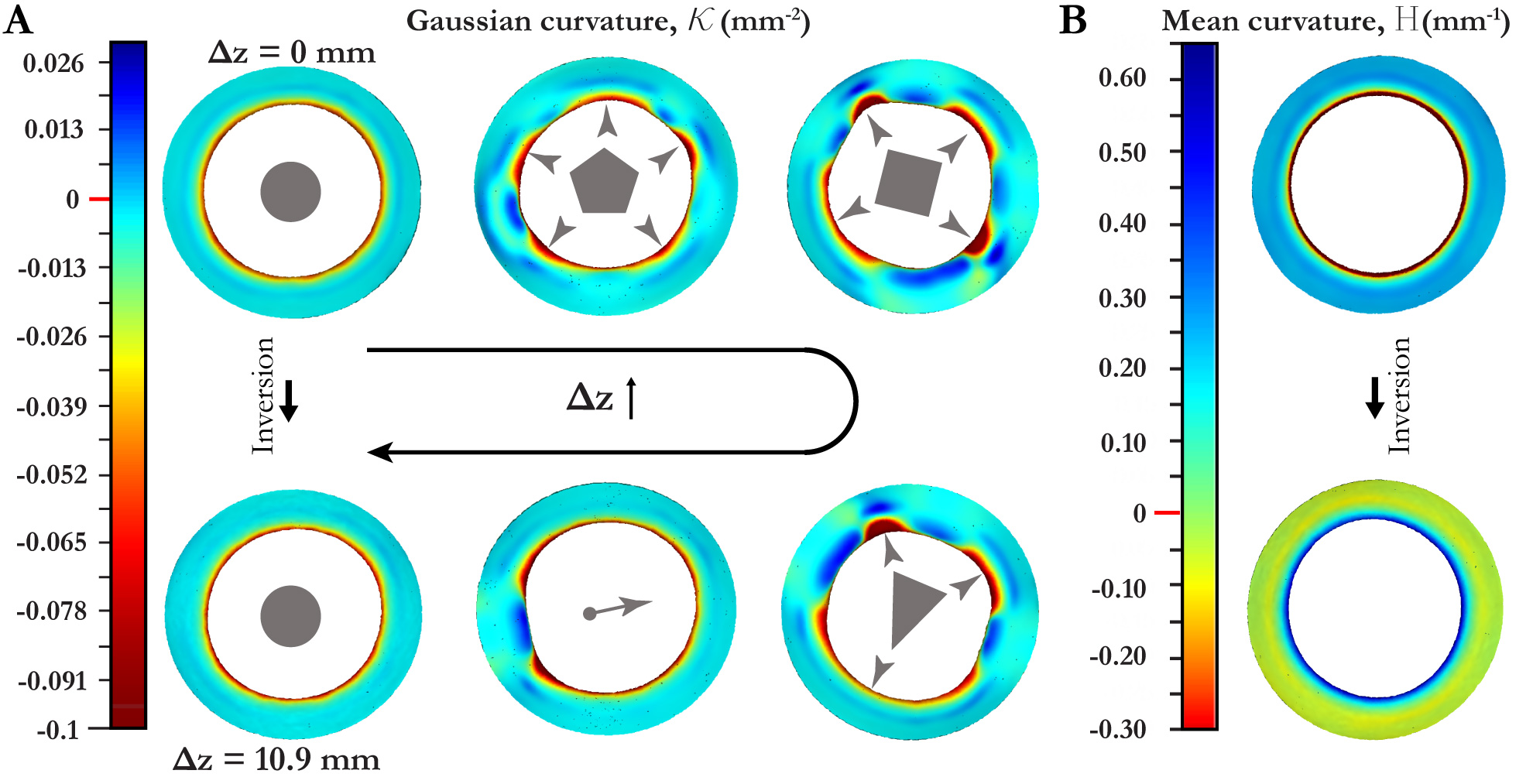}
    \caption
    {
    Distribution of Gaussian ($\mathcal{K}$) and mean ($H$) curvature during axial deformation of \abb with $h_1 \textrm{ = } 10 \textrm{ mm, } \alpha_1 \textrm{ = } 45^o \textrm{ and } O \textrm{ = } 0^o$:
    \textbf{(A)} Measured Gaussian curvature ($\mathcal{K}$) from X-ray computed tomography data, for axial indentation depths $\Delta z$. After an initial axisymmetric deformation regime, we observe periodic buckling of the central region upon axial indentation characterized with an initial mode of $m = 5$ at $\Delta z = 2.9$ mm, gradually depleting due to merging `crest' regions at $\Delta z = 7.9 \textrm{ mm } (m = 4) \textrm{, } 8.9 \textrm{ mm } (m = 3)$ and $10.8 \textrm{ mm } (m = 2)$, before regaining an axisymmetry in the isometric state at $\Delta z = 10.9$ mm. As clearly seen in color-maps, a majority of Gaussian curvature is concentrated in the central buckled region.
    \textbf{(B)} Measured mean curvature $H$ of isometric states at $\Delta z \textrm{ = } 0 \textrm{ and } 10.9$ mm.
    }
    \label{axialCT}
\end{figure*}

\section*{Four bar linkage model}
We further describe the 4-bar linkage model constructed as in the main text, Fig.5.  We can write the Lagrangian of this configuration as,
\vspace{-0.2em}
\begin{align}
  \mathcal{L}=&\frac{1}{2} K_r (\tau_l - \alpha_2) ^2 + \frac{1}{2} K_r (\pi - \tau_r - \alpha_2) ^2 +\nonumber\\
  &\lambda [(2R+W\cos \tau_r-W \cos \tau_l)^2+(W \sin \tau_r - W \sin \tau_l)^2\nonumber\\
  &-(2R-2W \cos \beta)^2]
  \label{Lagrangian_SI}
\end{align}
where the last term quadratically enforces the constraint that the two ends of the crank links without any springs are the same distance apart as the floating bar length. Upon variation of the angles $\tau_l$ and $\tau_r$, the resulting Euler-Lagrange equations are,
\begin{align}
  &K_r (\tau_l - \alpha_2) +2 \lambda [2RW \sin \tau_l +W^2 \sin(\tau_l - \tau_r)]=0 \label{Governequa1} \\
  &K_r (\tau_r + \alpha_2 - \pi) +2 \lambda [-2RW \sin \tau_r +W^2 \sin(\tau_r - \tau_l)]=0 \label{Governequa2}
\end{align}
Eq.\ref{Governequa1} and Eq.\ref{Governequa2} can be combined to eliminate the multiplier $\lambda$.  Dividing out the only modulus $K_r$ leaves us with one equilibrium equation and one constraint equation,
\begin{align}
  (\tau_l - \alpha_2)(W \sin (\tau_r -\tau_l)-2R \sin (\tau_r))+\nonumber\\(\tau_r-\pi+\alpha_2)(W \sin (\tau_r-\tau_l)-2R \sin \tau_l) =0 \label{Governequacombined}  \\
  (2R+W\cos \tau_r - W \cos \tau_l)^2+(W \sin \tau_r - W \sin \tau_l)^2-\nonumber\\(2R-2W \cos \beta)^2 =0 \label{constraintequa}
\end{align}

We numerically continue equilibria of these equations along deformation paths using AUTO 07P \cite{doedel2007auto}.  The behavior of the total energy of the system along these curves provides stability information.

Fig.\ref{energy_4bar} shows 4 closed curves corresponding to the energies of states represented by dots in Fig.5 (main text), with one link angle $\tau_l$ as parameter. 
For all of these curves, state \textit{(e)}, analogous to the extended state of a straw, is always a minimum on the lower right, but not always the ground state. 
Two equivalent-energy bent states \textit{(b)} appear on the green curve, where they are the ground states, and on the purple curve, where they are a little bit higher energy than the \textit{(e)} state.  On all curves, the fully inverted axially symmetric state \textit{(c)}, analogous to the collapsed state of a straw, is on the upper left. It is a maximum for the green and red curves and a minimum for the blue and purple curves.

We can move from the red to the green curve by decreasing the original cone angle $\alpha_2$ and increasing the mismatch $\beta - \alpha_2$, creating the two partially inverted local minima \textit{(b)}. From here we can create the collapsed local minimum \textit{(c)} by either further increasing the mismatch to move to the purple curve, or increasing the original cone angle to move to the blue curve; in the latter case, we lose the partially inverted states.
As can be seen from Fig.5 (main text), simply increasing mismatch at fixed original cone angle will eventually stabilize all the available states, but the order in which this happens depends on the original cone angle.

\section*{Understanding deformation of \abb through curvature analysis}

\noindent\textbf{\textit{Curvature measurement by in-situ X-ray CT: }}
\newline A Perkin Elmer IVIS SpectrumCT imager is used to obtain X-ray computed tomograms by exposing samples at a 23 mGy dose through a 120-mil-thick copper foil filter. The resulting scans (voxel size 150 $\mu$m) are thresholded accordingly in ImageJ for generating a 3D image stack for the desired geometry. To avoid possible interference during curvature measurement, the outer and inner surfaces of shell are separately identified using MATLAB (Mathworks Inc., USA), and a 3D point cloud depicting the inner surface is exported to the \text{Meshlab} software package \cite{meshlab_2008}. Therein, following a reduction of dense point clouds \cite{meshlab_PC_2012}, surface curvatures are computed using the algebraic point set surface method \cite{apss_2007}.

\vspace{1em}
\noindent\textbf{\textit{Curvature measurement during axial deformation of \abb: }}
\newline Curvature evolution during axial deformation is measured for \abb ($h_1 \textrm{ = } 10 \textrm{ mm, } \alpha_1 \textrm{ = } 45^o \textrm{ and } O \textrm{ = } 0^o$) indenting the shell quasi-statically with a step size of $<1$ mm. $\mathcal{K}$ is superimposed on the deformed geometries at various $\Delta z$ for analysis in Fig.\ref{axialCT}A as seen from below. At $\Delta z \textrm{ = } 0$, an expected  $\mathcal{K} \textrm{ = } 0$ is observed except in the central region, where $\mathcal{K}<0$. Upon further indentation, breaking of axisymmetry is observed, similar to the load-displacement experiments. As marked in Fig.\ref{axialCT}A, this buckled region appears to be periodic at $\Delta z \textrm{ = } 2.9$ mm, characterized by a periodicity $m \textrm{ = } 5$. Proceeding through the next steps of indentation at $\Delta z \textrm{ = } 4 \textrm{, } 8.9 \textrm{ and } 10.8$ mm, we consistently note two `crest' regions merging into one, changing the mode from $m \textrm{ = } 5$, to $m \textrm{ = } 4 \textrm{, } 3 \textrm{ and } 2$ along the each step. A stable full inversion of the lower frustum restores axial symmetry and distribution of $\mathcal{K}$ at $\Delta z \textrm{ = } 10.9$ mm. $H$ of isometric states ($\Delta z \textrm{ = } 0 \textrm{ and }10.9$) is also plotted in Fig.\ref{axialCT}B, demonstrating the expected inversion in sign.

\section*{Quantifying overcurvature introduced in 3D printed \abb}

\begin{figure*}[t]
    \centering
    \vspace{0.8em}
    \includegraphics[width=0.9\textwidth]{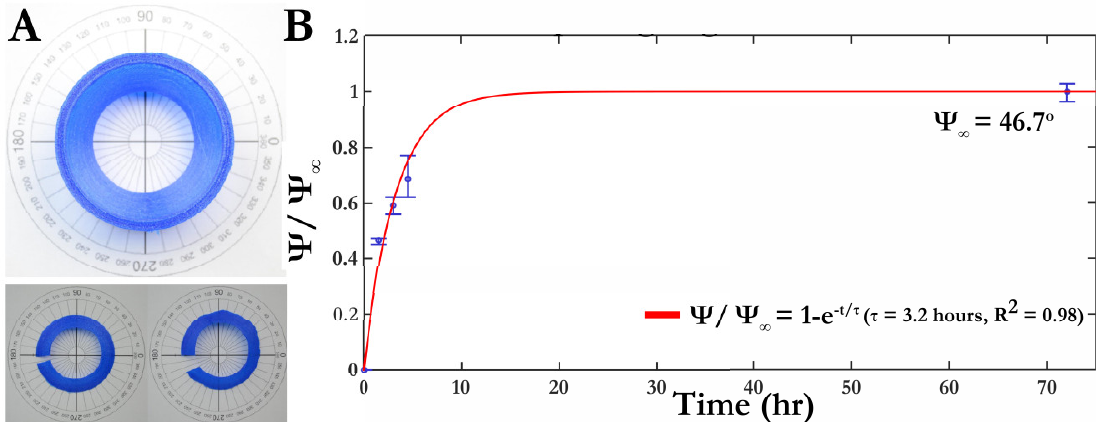}
    \caption
    {Opening angle $\psi$ (as proxy of overcurvature) of 3D printed \abb as a function of time spent in an axially collapsed state.
    }
    \label{3dstress}
\end{figure*}

Controlled amounts of overcurvature in \abb fabricated through direct 3D printing of poly(urethane) can be introduced as follows. After being printed in an extended state, the \abb are transferred into a harness that holds the structure in an axially collapsed position for extended periods of time. After time $t$, samples are removed from the harness, examined for multistability and cut open. Similar to \textit{Pop toobs}, the 3D printed \abb also open radially upon relaxation (Fig.\ref{3dstress}). We measure the opening angle $\psi$, and repeat this experiment for samples that are constrained for different time periods ($t$). Fig.\ref{3dstress}B shows the dependence of opening angle normalized by the value at $t = 72$ ($\psi_\infty$), along with a one-parameter exponential fit. For the poly(urethane) in this study, keeping \abb ($h_1 = 10$ mm, $\alpha_1 = 45^o$) in collapsed state for approximately 3 h is enough to induce sufficient overcurvature for stability of the bent state, closely matching the time-scale over which pre-stress is developed in Fig. \ref{3dstress}B.

\section*{{\huge References}}


%


\clearpage

\hypertarget{SI_M}{}

\renewcommand{\thevideo}{M\arabic{video}}

\begin{video*} 
    \includegraphics[width=0.6\textwidth]{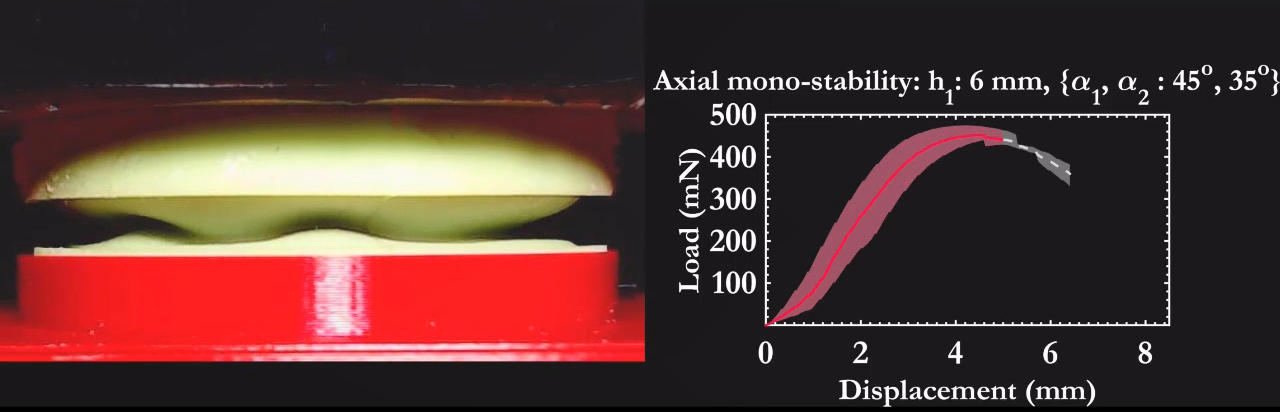}
    \label{M1}
    \setfloatlink{https://streamable.com/sk9sm}
    \caption{Load-displacement curves and corresponding images showing deformation of \abb under axial loading. Samples lacking ($h_1 = 6$ mm), and exhibiting ($h_1 = 10$ mm) axial bistability are shown. \url{https://streamable.com/sk9sm}} 
\end{video*} 

\begin{video*} 
    \includegraphics[width=0.5\textwidth]{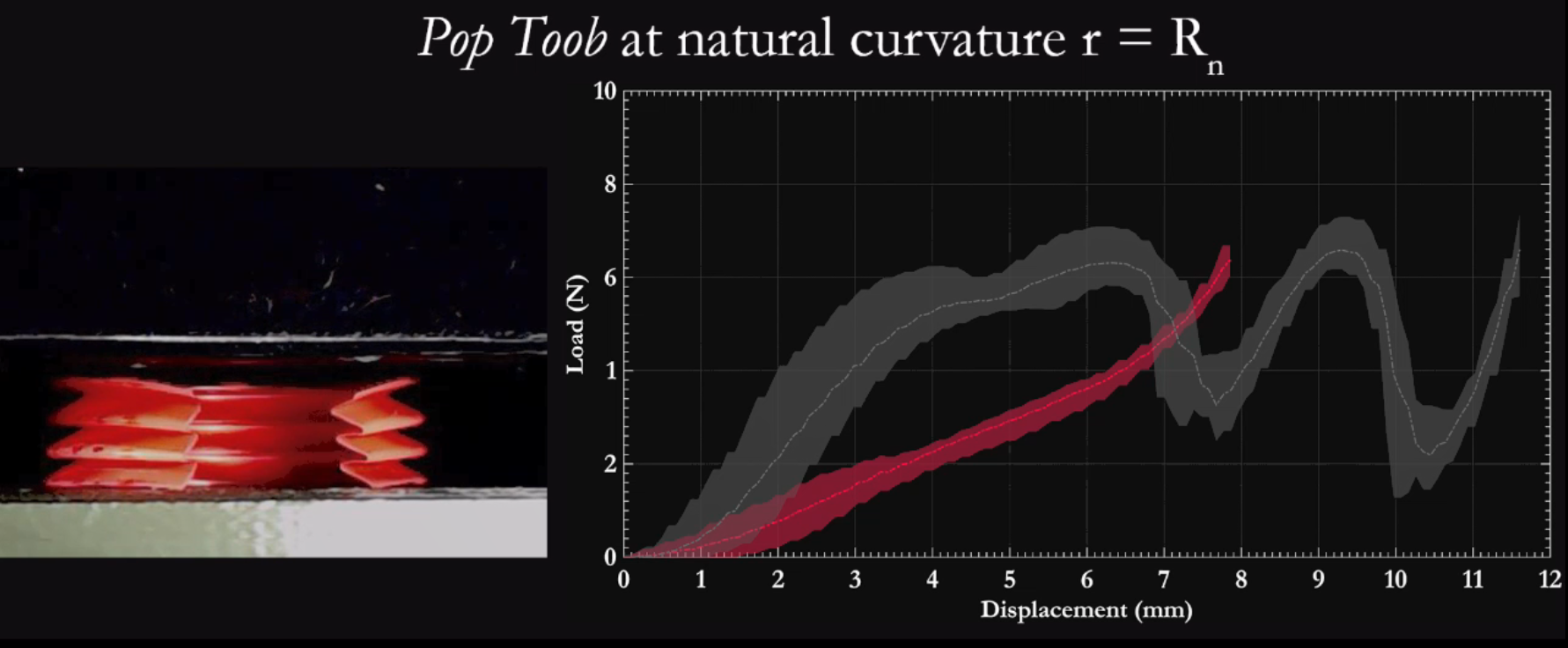}
    \label{M2}
    \setfloatlink{https://streamable.com/fovo7}
    \caption{Load-displacement curves and corresponding images showing axial deformation of a \textit{Pop Toob} with its natural curvature, exhibiting smooth accordian-like deformation, and in the overcurved state, exhibiting multistability. \url{https://streamable.com/fovo7}} 
\end{video*} 

\begin{video*} 
    \includegraphics[width=0.6\textwidth]{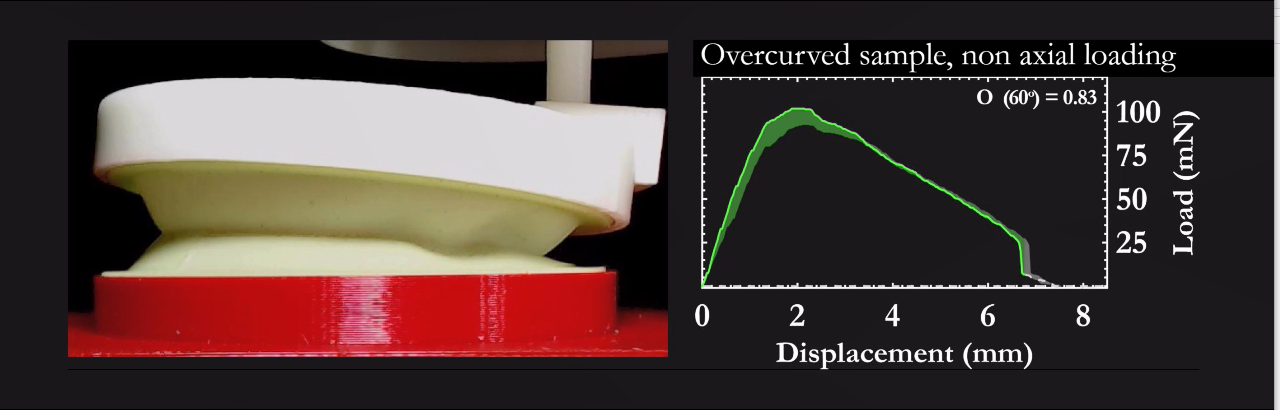}
    \label{M3}
    \setfloatlink{https://streamable.com/pr3sk}
    \caption{Load-displacement curves and corresponding images showing non-axial deformation of control ($O = 0$) and overcurved ($O = 0.17$) \abb with $h_1 = 6$ mm, $\alpha_1 = 45^o$ and $\alpha_2 = 35^o$. The control sample lacks stability in the bent state and shows a smaller number and amplitude of wrinkles in the transition state.\url{https://streamable.com/pr3sk}} 
\end{video*} 

\begin{video*} 
    \includegraphics[width=0.4\textwidth]{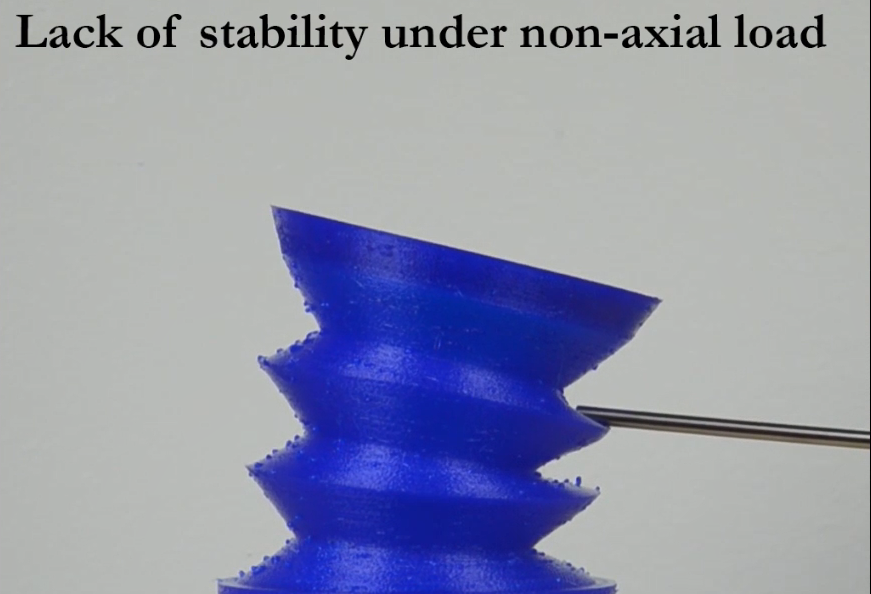}
    \label{M4}
    \setfloatlink{https://streamable.com/o63bi}
    \caption{Images showing axial and non-axial loading of 3D printed \abb, revealing a change in stability due to the introduction of overcurvature. \url{https://streamable.com/o63bi}} 
\end{video*} 


\begin{thebibliography}{40}%
\makeatletter
\providecommand \@ifxundefined [1]{%
 \@ifx{#1\undefined}
}%
\providecommand \@ifnum [1]{%
 \ifnum #1\expandafter \@firstoftwo
 \else \expandafter \@secondoftwo
 \fi
}%
\providecommand \@ifx [1]{%
 \ifx #1\expandafter \@firstoftwo
 \else \expandafter \@secondoftwo
 \fi
}%
\providecommand \natexlab [1]{#1}%
\providecommand \enquote  [1]{``#1''}%
\providecommand \bibnamefont  [1]{#1}%
\providecommand \bibfnamefont [1]{#1}%
\providecommand \citenamefont [1]{#1}%
\providecommand \href@noop [0]{\@secondoftwo}%
\providecommand \href [0]{\begingroup \@sanitize@url \@href}%
\providecommand \@href[1]{\@@startlink{#1}\@@href}%
\providecommand \@@href[1]{\endgroup#1\@@endlink}%
\providecommand \@sanitize@url [0]{\catcode `\\12\catcode `\$12\catcode
  `\&12\catcode `\#12\catcode `\^12\catcode `\_12\catcode `\%12\relax}%
\providecommand \@@startlink[1]{}%
\providecommand \@@endlink[0]{}%
\providecommand \url  [0]{\begingroup\@sanitize@url \@url }%
\providecommand \@url [1]{\endgroup\@href {#1}{\urlprefix }}%
\providecommand \urlprefix  [0]{URL }%
\providecommand \Eprint [0]{\href }%
\providecommand \doibase [0]{http://dx.doi.org/}%
\providecommand \selectlanguage [0]{\@gobble}%
\providecommand \bibinfo  [0]{\@secondoftwo}%
\providecommand \bibfield  [0]{\@secondoftwo}%
\providecommand \translation [1]{[#1]}%
\providecommand \BibitemOpen [0]{}%
\providecommand \bibitemStop [0]{}%
\providecommand \bibitemNoStop [0]{.\EOS\space}%
\providecommand \EOS [0]{\spacefactor3000\relax}%
\providecommand \BibitemShut  [1]{\csname bibitem#1\endcsname}%
\let\auto@bib@innerbib\@empty
\bibitem [{\citenamefont {Hu}\ and\ \citenamefont
  {Burgue{\~n}o}(2015)}]{hu2015buckling}%
  \BibitemOpen
  \bibfield  {author} {\bibinfo {author} {\bibfnamefont {N.}~\bibnamefont
  {Hu}}\ and\ \bibinfo {author} {\bibfnamefont {R.}~\bibnamefont
  {Burgue{\~n}o}},\ }\href@noop {} {\bibfield  {journal} {\bibinfo  {journal}
  {Smart Materials and Structures}\ }\textbf {\bibinfo {volume} {24}},\
  \bibinfo {pages} {063001} (\bibinfo {year} {2015})}\BibitemShut {NoStop}%
\bibitem [{\citenamefont {Calladine}(1989)}]{calladine1989theory}%
  \BibitemOpen
  \bibfield  {author} {\bibinfo {author} {\bibfnamefont {C.~R.}\ \bibnamefont
  {Calladine}},\ }\href@noop {} {\emph {\bibinfo {title} {Theory of shell
  structures}}}\ (\bibinfo  {publisher} {Cambridge University Press},\ \bibinfo
  {year} {1989})\BibitemShut {NoStop}%
\bibitem [{\citenamefont {Pippard}(1990)}]{pippard1990elastic}%
  \BibitemOpen
  \bibfield  {author} {\bibinfo {author} {\bibfnamefont {A.}~\bibnamefont
  {Pippard}},\ }\href@noop {} {\bibfield  {journal} {\bibinfo  {journal}
  {European Journal of Physics}\ }\textbf {\bibinfo {volume} {11}},\ \bibinfo
  {pages} {359} (\bibinfo {year} {1990})}\BibitemShut {NoStop}%
\bibitem [{\citenamefont {Seffen}(2006)}]{seffen2006metal}%
  \BibitemOpen
  \bibfield  {author} {\bibinfo {author} {\bibfnamefont {K.}~\bibnamefont
  {Seffen}},\ }\href@noop {} {\bibfield  {journal} {\bibinfo  {journal}
  {Scripta materialia}\ }\textbf {\bibinfo {volume} {55}},\ \bibinfo {pages}
  {411} (\bibinfo {year} {2006})}\BibitemShut {NoStop}%
\bibitem [{\citenamefont {Pandey}\ \emph {et~al.}(2014)\citenamefont {Pandey},
  \citenamefont {Moulton}, \citenamefont {Vella},\ and\ \citenamefont
  {Holmes}}]{pandey2014}%
  \BibitemOpen
  \bibfield  {author} {\bibinfo {author} {\bibfnamefont {A.}~\bibnamefont
  {Pandey}}, \bibinfo {author} {\bibfnamefont {D.~E.}\ \bibnamefont {Moulton}},
  \bibinfo {author} {\bibfnamefont {D.}~\bibnamefont {Vella}}, \ and\ \bibinfo
  {author} {\bibfnamefont {D.~P.}\ \bibnamefont {Holmes}},\ }\href@noop {}
  {\bibfield  {journal} {\bibinfo  {journal} {Europhys. Lett.}\ }\textbf
  {\bibinfo {volume} {105}},\ \bibinfo {pages} {24001} (\bibinfo {year}
  {2014})}\BibitemShut {NoStop}%
\bibitem [{\citenamefont {Reid}\ \emph {et~al.}(2017)\citenamefont {Reid},
  \citenamefont {Lechenault}, \citenamefont {Rica},\ and\ \citenamefont
  {Adda-Bedia}}]{reid2017geometry}%
  \BibitemOpen
  \bibfield  {author} {\bibinfo {author} {\bibfnamefont {A.}~\bibnamefont
  {Reid}}, \bibinfo {author} {\bibfnamefont {F.}~\bibnamefont {Lechenault}},
  \bibinfo {author} {\bibfnamefont {S.}~\bibnamefont {Rica}}, \ and\ \bibinfo
  {author} {\bibfnamefont {M.}~\bibnamefont {Adda-Bedia}},\ }\href@noop {}
  {\bibfield  {journal} {\bibinfo  {journal} {Physical Review E}\ }\textbf
  {\bibinfo {volume} {95}},\ \bibinfo {pages} {013002} (\bibinfo {year}
  {2017})}\BibitemShut {NoStop}%
\bibitem [{\citenamefont {Holmes}\ and\ \citenamefont
  {Crosby}(2007)}]{crosby2007lens}%
  \BibitemOpen
  \bibfield  {author} {\bibinfo {author} {\bibfnamefont {D.}~\bibnamefont
  {Holmes}}\ and\ \bibinfo {author} {\bibfnamefont {A.}~\bibnamefont
  {Crosby}},\ }\href {\doibase 10.1002/adma.200700584} {\bibfield  {journal}
  {\bibinfo  {journal} {Adv. Mater.}\ }\textbf {\bibinfo {volume} {19}},\
  \bibinfo {pages} {3589} (\bibinfo {year} {2007})}\BibitemShut {NoStop}%
\bibitem [{\citenamefont {Tavakol}\ \emph {et~al.}(2014)\citenamefont
  {Tavakol}, \citenamefont {Bozlar}, \citenamefont {Punckt}, \citenamefont
  {Froehlicher}, \citenamefont {Stone}, \citenamefont {Aksay},\ and\
  \citenamefont {Holmes}}]{holmes2014pump}%
  \BibitemOpen
  \bibfield  {author} {\bibinfo {author} {\bibfnamefont {B.}~\bibnamefont
  {Tavakol}}, \bibinfo {author} {\bibfnamefont {M.}~\bibnamefont {Bozlar}},
  \bibinfo {author} {\bibfnamefont {C.}~\bibnamefont {Punckt}}, \bibinfo
  {author} {\bibfnamefont {G.}~\bibnamefont {Froehlicher}}, \bibinfo {author}
  {\bibfnamefont {H.~A.}\ \bibnamefont {Stone}}, \bibinfo {author}
  {\bibfnamefont {I.~A.}\ \bibnamefont {Aksay}}, \ and\ \bibinfo {author}
  {\bibfnamefont {D.~P.}\ \bibnamefont {Holmes}},\ }\href@noop {} {\bibfield
  {journal} {\bibinfo  {journal} {Soft Matter}\ }\textbf {\bibinfo {volume}
  {10}},\ \bibinfo {pages} {4789} (\bibinfo {year} {2014})}\BibitemShut
  {NoStop}%
\bibitem [{\citenamefont {Shankar}\ \emph {et~al.}(2013)\citenamefont
  {Shankar}, \citenamefont {Smith}, \citenamefont {Tondiglia}, \citenamefont
  {Lee}, \citenamefont {McConney}, \citenamefont {Wang}, \citenamefont {Tan},\
  and\ \citenamefont {White}}]{shankar2013contactless}%
  \BibitemOpen
  \bibfield  {author} {\bibinfo {author} {\bibfnamefont {M.~R.}\ \bibnamefont
  {Shankar}}, \bibinfo {author} {\bibfnamefont {M.~L.}\ \bibnamefont {Smith}},
  \bibinfo {author} {\bibfnamefont {V.~P.}\ \bibnamefont {Tondiglia}}, \bibinfo
  {author} {\bibfnamefont {K.~M.}\ \bibnamefont {Lee}}, \bibinfo {author}
  {\bibfnamefont {M.~E.}\ \bibnamefont {McConney}}, \bibinfo {author}
  {\bibfnamefont {D.~H.}\ \bibnamefont {Wang}}, \bibinfo {author}
  {\bibfnamefont {L.-S.}\ \bibnamefont {Tan}}, \ and\ \bibinfo {author}
  {\bibfnamefont {T.~J.}\ \bibnamefont {White}},\ }\href@noop {} {\bibfield
  {journal} {\bibinfo  {journal} {Proc. Natl. Acad. Sci. U.S.A.}\ }\textbf
  {\bibinfo {volume} {110}},\ \bibinfo {pages} {18792} (\bibinfo {year}
  {2013})}\BibitemShut {NoStop}%
\bibitem [{\citenamefont {Ramachandran}\ \emph {et~al.}(2016)\citenamefont
  {Ramachandran}, \citenamefont {Bartlett}, \citenamefont {Wissman},\ and\
  \citenamefont {Majidi}}]{bartlett2016elastic}%
  \BibitemOpen
  \bibfield  {author} {\bibinfo {author} {\bibfnamefont {V.}~\bibnamefont
  {Ramachandran}}, \bibinfo {author} {\bibfnamefont {M.~D.}\ \bibnamefont
  {Bartlett}}, \bibinfo {author} {\bibfnamefont {J.}~\bibnamefont {Wissman}}, \
  and\ \bibinfo {author} {\bibfnamefont {C.}~\bibnamefont {Majidi}},\
  }\href@noop {} {\bibfield  {journal} {\bibinfo  {journal} {Extreme Mechanics
  Letters}\ }\textbf {\bibinfo {volume} {9}},\ \bibinfo {pages} {282} (\bibinfo
  {year} {2016})}\BibitemShut {NoStop}%
\bibitem [{\citenamefont {Zirbel}\ \emph {et~al.}(2016)\citenamefont {Zirbel},
  \citenamefont {Tolman}, \citenamefont {Trease},\ and\ \citenamefont
  {Howell}}]{zirbel2016bistable}%
  \BibitemOpen
  \bibfield  {author} {\bibinfo {author} {\bibfnamefont {S.~A.}\ \bibnamefont
  {Zirbel}}, \bibinfo {author} {\bibfnamefont {K.~A.}\ \bibnamefont {Tolman}},
  \bibinfo {author} {\bibfnamefont {B.~P.}\ \bibnamefont {Trease}}, \ and\
  \bibinfo {author} {\bibfnamefont {L.~L.}\ \bibnamefont {Howell}},\
  }\href@noop {} {\bibfield  {journal} {\bibinfo  {journal} {PloS one}\
  }\textbf {\bibinfo {volume} {11}},\ \bibinfo {pages} {e0168218} (\bibinfo
  {year} {2016})}\BibitemShut {NoStop}%
\bibitem [{\citenamefont {Forterre}\ \emph {et~al.}(2005)\citenamefont
  {Forterre}, \citenamefont {Skotheim}, \citenamefont {Dumais},\ and\
  \citenamefont {Mahadevan}}]{forterre2005venus}%
  \BibitemOpen
  \bibfield  {author} {\bibinfo {author} {\bibfnamefont {Y.}~\bibnamefont
  {Forterre}}, \bibinfo {author} {\bibfnamefont {J.~M.}\ \bibnamefont
  {Skotheim}}, \bibinfo {author} {\bibfnamefont {J.}~\bibnamefont {Dumais}}, \
  and\ \bibinfo {author} {\bibfnamefont {L.}~\bibnamefont {Mahadevan}},\
  }\href@noop {} {\bibfield  {journal} {\bibinfo  {journal} {Nature}\ }\textbf
  {\bibinfo {volume} {433}},\ \bibinfo {pages} {421} (\bibinfo {year}
  {2005})}\BibitemShut {NoStop}%
\bibitem [{\citenamefont {Hayashi}\ \emph {et~al.}(2009)\citenamefont
  {Hayashi}, \citenamefont {Feilich},\ and\ \citenamefont
  {Ellerby}}]{hayashi2009mechanics}%
  \BibitemOpen
  \bibfield  {author} {\bibinfo {author} {\bibfnamefont {M.}~\bibnamefont
  {Hayashi}}, \bibinfo {author} {\bibfnamefont {K.~L.}\ \bibnamefont
  {Feilich}}, \ and\ \bibinfo {author} {\bibfnamefont {D.~J.}\ \bibnamefont
  {Ellerby}},\ }\href {\doibase 10.1093/jxb/erp070} {\bibfield  {journal}
  {\bibinfo  {journal} {Journal of Experimental Botany}\ }\textbf {\bibinfo
  {volume} {60}},\ \bibinfo {pages} {2045} (\bibinfo {year}
  {2009})}\BibitemShut {NoStop}%
\bibitem [{\citenamefont {Smith}\ \emph {et~al.}(2011)\citenamefont {Smith},
  \citenamefont {Yanega},\ and\ \citenamefont {Ruina}}]{hummingbird2011}%
  \BibitemOpen
  \bibfield  {author} {\bibinfo {author} {\bibfnamefont {M.}~\bibnamefont
  {Smith}}, \bibinfo {author} {\bibfnamefont {G.}~\bibnamefont {Yanega}}, \
  and\ \bibinfo {author} {\bibfnamefont {A.}~\bibnamefont {Ruina}},\
  }\href@noop {} {\bibfield  {journal} {\bibinfo  {journal} {J. Theor. Biol.}\
  }\textbf {\bibinfo {volume} {282}},\ \bibinfo {pages} {41} (\bibinfo {year}
  {2011})}\BibitemShut {NoStop}%
\bibitem [{\citenamefont {Son}\ \emph {et~al.}(2013)\citenamefont {Son},
  \citenamefont {Guasto},\ and\ \citenamefont {Stocker}}]{2013bacteria}%
  \BibitemOpen
  \bibfield  {author} {\bibinfo {author} {\bibfnamefont {K.}~\bibnamefont
  {Son}}, \bibinfo {author} {\bibfnamefont {J.~S.}\ \bibnamefont {Guasto}}, \
  and\ \bibinfo {author} {\bibfnamefont {R.}~\bibnamefont {Stocker}},\
  }\href@noop {} {\bibfield  {journal} {\bibinfo  {journal} {Nature Physics}\
  }\textbf {\bibinfo {volume} {9}},\ \bibinfo {pages} {494} (\bibinfo {year}
  {2013})}\BibitemShut {NoStop}%
\bibitem [{\citenamefont {Turk}\ and\ \citenamefont
  {Levoy}(1994)}]{stanford_bunny1994}%
  \BibitemOpen
  \bibfield  {author} {\bibinfo {author} {\bibfnamefont {G.}~\bibnamefont
  {Turk}}\ and\ \bibinfo {author} {\bibfnamefont {M.}~\bibnamefont {Levoy}},\
  }in\ \href@noop {} {\emph {\bibinfo {booktitle} {Proceedings of the 21st
  annual conference on Computer graphics and interactive techniques}}}\
  (\bibinfo {organization} {ACM},\ \bibinfo {year} {1994})\ pp.\ \bibinfo
  {pages} {311--318}\BibitemShut {NoStop}%
\bibitem [{\citenamefont {Friedman}(1951)}]{friedman1951flexible}%
  \BibitemOpen
  \bibfield  {author} {\bibinfo {author} {\bibfnamefont {J.~B.}\ \bibnamefont
  {Friedman}},\ }\href@noop {} {\enquote {\bibinfo {title} {Flexible drinking
  straw},}\ } (\bibinfo {year} {1951}),\ \bibinfo {note} {uS Patent
  2,550,797}\BibitemShut {NoStop}%
\bibitem [{\citenamefont {Harp}\ \emph {et~al.}(1968)\citenamefont {Harp},
  \citenamefont {Leible},\ and\ \citenamefont {Mccort}}]{harp1968}%
  \BibitemOpen
  \bibfield  {author} {\bibinfo {author} {\bibfnamefont {H.~J.}\ \bibnamefont
  {Harp}}, \bibinfo {author} {\bibfnamefont {W.~T.}\ \bibnamefont {Leible}}, \
  and\ \bibinfo {author} {\bibfnamefont {W.~M.}\ \bibnamefont {Mccort}},\
  }\href@noop {} {\enquote {\bibinfo {title} {Flexible drinking tube},}\ }
  (\bibinfo {year} {1968}),\ \bibinfo {note} {uS Patent 3,409,224}\BibitemShut
  {NoStop}%
\bibitem [{\citenamefont {Mikol}(1989)}]{mikol1989adjustable}%
  \BibitemOpen
  \bibfield  {author} {\bibinfo {author} {\bibfnamefont {E.}~\bibnamefont
  {Mikol}},\ }\href@noop {} {\enquote {\bibinfo {title} {Adjustable tubular
  wall structure for connectors and the like},}\ } (\bibinfo {year} {1989}),\
  \bibinfo {note} {uS Patent 4,846,510}\BibitemShut {NoStop}%
\bibitem [{\citenamefont {Diebolt}\ and\ \citenamefont
  {Hendrickson}(1975)}]{1975tubular}%
  \BibitemOpen
  \bibfield  {author} {\bibinfo {author} {\bibfnamefont {E.~J.}\ \bibnamefont
  {Diebolt}}\ and\ \bibinfo {author} {\bibfnamefont {R.~A.}\ \bibnamefont
  {Hendrickson}},\ }\href@noop {} {\enquote {\bibinfo {title} {Tubular hinge
  assembly},}\ } (\bibinfo {year} {1975}),\ \bibinfo {note} {uS Patent
  3,929,165}\BibitemShut {NoStop}%
\bibitem [{\citenamefont {Kusuma}\ \emph {et~al.}(2010)\citenamefont {Kusuma},
  \citenamefont {Card},\ and\ \citenamefont {Lugo}}]{2010collapsible}%
  \BibitemOpen
  \bibfield  {author} {\bibinfo {author} {\bibfnamefont {D.}~\bibnamefont
  {Kusuma}}, \bibinfo {author} {\bibfnamefont {P.~M.}\ \bibnamefont {Card}}, \
  and\ \bibinfo {author} {\bibfnamefont {H.~J.~B.}\ \bibnamefont {Lugo}},\
  }\href@noop {} {\enquote {\bibinfo {title} {Collapsible container},}\ }
  (\bibinfo {year} {2010}),\ \bibinfo {note} {uS Patent 7,654,402}\BibitemShut
  {NoStop}%
\bibitem [{\citenamefont {Bende}\ \emph {et~al.}(2015)\citenamefont {Bende},
  \citenamefont {Evans}, \citenamefont {Innes-Gold}, \citenamefont {Marin},
  \citenamefont {Cohen}, \citenamefont {Hayward},\ and\ \citenamefont
  {Santangelo}}]{bende2015snap}%
  \BibitemOpen
  \bibfield  {author} {\bibinfo {author} {\bibfnamefont {N.~P.}\ \bibnamefont
  {Bende}}, \bibinfo {author} {\bibfnamefont {A.~A.}\ \bibnamefont {Evans}},
  \bibinfo {author} {\bibfnamefont {S.}~\bibnamefont {Innes-Gold}}, \bibinfo
  {author} {\bibfnamefont {L.~A.}\ \bibnamefont {Marin}}, \bibinfo {author}
  {\bibfnamefont {I.}~\bibnamefont {Cohen}}, \bibinfo {author} {\bibfnamefont
  {R.~C.}\ \bibnamefont {Hayward}}, \ and\ \bibinfo {author} {\bibfnamefont
  {C.~D.}\ \bibnamefont {Santangelo}},\ }\href@noop {} {\bibfield  {journal}
  {\bibinfo  {journal} {Proceedings of the National Academy of Sciences}\
  }\textbf {\bibinfo {volume} {112}},\ \bibinfo {pages} {11175} (\bibinfo
  {year} {2015})}\BibitemShut {NoStop}%
\bibitem [{\citenamefont {Mouthuy}\ \emph {et~al.}(2012)\citenamefont
  {Mouthuy}, \citenamefont {Coulombier}, \citenamefont {Pardoen}, \citenamefont
  {Raskin},\ and\ \citenamefont {Jonas}}]{mouthuy2012overcurvature}%
  \BibitemOpen
  \bibfield  {author} {\bibinfo {author} {\bibfnamefont {P.-O.}\ \bibnamefont
  {Mouthuy}}, \bibinfo {author} {\bibfnamefont {M.}~\bibnamefont {Coulombier}},
  \bibinfo {author} {\bibfnamefont {T.}~\bibnamefont {Pardoen}}, \bibinfo
  {author} {\bibfnamefont {J.-P.}\ \bibnamefont {Raskin}}, \ and\ \bibinfo
  {author} {\bibfnamefont {A.~M.}\ \bibnamefont {Jonas}},\ }\href@noop {}
  {\bibfield  {journal} {\bibinfo  {journal} {Nature communications}\ }\textbf
  {\bibinfo {volume} {3}},\ \bibinfo {pages} {1290} (\bibinfo {year}
  {2012})}\BibitemShut {NoStop}%
\bibitem [{\citenamefont {Dias}\ and\ \citenamefont
  {Audoly}(2015)}]{dias2015wunderlich}%
  \BibitemOpen
  \bibfield  {author} {\bibinfo {author} {\bibfnamefont {M.~A.}\ \bibnamefont
  {Dias}}\ and\ \bibinfo {author} {\bibfnamefont {B.}~\bibnamefont {Audoly}},\
  }\href@noop {} {\bibfield  {journal} {\bibinfo  {journal} {Journal of
  Elasticity}\ }\textbf {\bibinfo {volume} {119}},\ \bibinfo {pages} {49}
  (\bibinfo {year} {2015})}\BibitemShut {NoStop}%
\bibitem [{\citenamefont {Audoly}\ and\ \citenamefont
  {Seffen}(2015)}]{audoly2015buckling}%
  \BibitemOpen
  \bibfield  {author} {\bibinfo {author} {\bibfnamefont {B.}~\bibnamefont
  {Audoly}}\ and\ \bibinfo {author} {\bibfnamefont {K.~A.}\ \bibnamefont
  {Seffen}},\ }\href@noop {} {\bibfield  {journal} {\bibinfo  {journal}
  {Journal of Elasticity}\ }\textbf {\bibinfo {volume} {119}},\ \bibinfo
  {pages} {293} (\bibinfo {year} {2015})}\BibitemShut {NoStop}%
\bibitem [{\citenamefont {Efrati}\ \emph {et~al.}(2009)\citenamefont {Efrati},
  \citenamefont {Sharon},\ and\ \citenamefont {Kupferman}}]{Efrati09JMPS}%
  \BibitemOpen
  \bibfield  {author} {\bibinfo {author} {\bibfnamefont {E.}~\bibnamefont
  {Efrati}}, \bibinfo {author} {\bibfnamefont {E.}~\bibnamefont {Sharon}}, \
  and\ \bibinfo {author} {\bibfnamefont {R.}~\bibnamefont {Kupferman}},\
  }\href@noop {} {\bibfield  {journal} {\bibinfo  {journal} {J. Mech. Phys.
  Sol.}\ }\textbf {\bibinfo {volume} {57}},\ \bibinfo {pages} {762} (\bibinfo
  {year} {2009})}\BibitemShut {NoStop}%
\bibitem [{\citenamefont {Kebadze}\ \emph {et~al.}(2004)\citenamefont
  {Kebadze}, \citenamefont {Guest},\ and\ \citenamefont
  {Pellegrino}}]{kebadze2004bistable}%
  \BibitemOpen
  \bibfield  {author} {\bibinfo {author} {\bibfnamefont {E.}~\bibnamefont
  {Kebadze}}, \bibinfo {author} {\bibfnamefont {S.}~\bibnamefont {Guest}}, \
  and\ \bibinfo {author} {\bibfnamefont {S.}~\bibnamefont {Pellegrino}},\
  }\href@noop {} {\bibfield  {journal} {\bibinfo  {journal} {International
  Journal of Solids and Structures}\ }\textbf {\bibinfo {volume} {41}},\
  \bibinfo {pages} {2801} (\bibinfo {year} {2004})}\BibitemShut {NoStop}%
\bibitem [{\citenamefont {Seffen}\ and\ \citenamefont
  {Guest}(2011)}]{seffen2011prestressed}%
  \BibitemOpen
  \bibfield  {author} {\bibinfo {author} {\bibfnamefont {K.~A.}\ \bibnamefont
  {Seffen}}\ and\ \bibinfo {author} {\bibfnamefont {S.~D.}\ \bibnamefont
  {Guest}},\ }\href@noop {} {\bibfield  {journal} {\bibinfo  {journal} {Journal
  of Applied Mechanics}\ }\textbf {\bibinfo {volume} {78}},\ \bibinfo {pages}
  {011002} (\bibinfo {year} {2011})}\BibitemShut {NoStop}%
\bibitem [{\citenamefont {Hamouche}\ \emph {et~al.}(2016)\citenamefont
  {Hamouche}, \citenamefont {Maurini}, \citenamefont {Vincenti},\ and\
  \citenamefont {Vidoli}}]{hamouche2016basic}%
  \BibitemOpen
  \bibfield  {author} {\bibinfo {author} {\bibfnamefont {W.}~\bibnamefont
  {Hamouche}}, \bibinfo {author} {\bibfnamefont {C.}~\bibnamefont {Maurini}},
  \bibinfo {author} {\bibfnamefont {A.}~\bibnamefont {Vincenti}}, \ and\
  \bibinfo {author} {\bibfnamefont {S.}~\bibnamefont {Vidoli}},\ }\href@noop {}
  {\bibfield  {journal} {\bibinfo  {journal} {Meccanica}\ }\textbf {\bibinfo
  {volume} {51}},\ \bibinfo {pages} {2305} (\bibinfo {year}
  {2016})}\BibitemShut {NoStop}%
\bibitem [{\citenamefont {Moessner}\ and\ \citenamefont
  {Ramirez}(2006)}]{isling_model_2006}%
  \BibitemOpen
  \bibfield  {author} {\bibinfo {author} {\bibfnamefont {R.}~\bibnamefont
  {Moessner}}\ and\ \bibinfo {author} {\bibfnamefont {A.~P.}\ \bibnamefont
  {Ramirez}},\ }\href@noop {} {\bibfield  {journal} {\bibinfo  {journal} {Phys.
  Today}\ }\textbf {\bibinfo {volume} {59}},\ \bibinfo {pages} {24} (\bibinfo
  {year} {2006})}\BibitemShut {NoStop}%
\bibitem [{\citenamefont {Jensen}\ and\ \citenamefont
  {Howell}(2003)}]{jensen2003identification}%
  \BibitemOpen
  \bibfield  {author} {\bibinfo {author} {\bibfnamefont {B.~D.}\ \bibnamefont
  {Jensen}}\ and\ \bibinfo {author} {\bibfnamefont {L.~L.}\ \bibnamefont
  {Howell}},\ }\href@noop {} {\bibfield  {journal} {\bibinfo  {journal}
  {Journal of Mechanical Design}\ }\textbf {\bibinfo {volume} {125}},\ \bibinfo
  {pages} {701} (\bibinfo {year} {2003})}\BibitemShut {NoStop}%
\bibitem [{\citenamefont {Jensen}\ \emph {et~al.}(1999)\citenamefont {Jensen},
  \citenamefont {Howell},\ and\ \citenamefont {Salmon}}]{jensen1999design}%
  \BibitemOpen
  \bibfield  {author} {\bibinfo {author} {\bibfnamefont {B.}~\bibnamefont
  {Jensen}}, \bibinfo {author} {\bibfnamefont {L.}~\bibnamefont {Howell}}, \
  and\ \bibinfo {author} {\bibfnamefont {L.}~\bibnamefont {Salmon}},\
  }\href@noop {} {\bibfield  {journal} {\bibinfo  {journal} {Journal of
  Mechanical Design}\ }\textbf {\bibinfo {volume} {121}},\ \bibinfo {pages}
  {416} (\bibinfo {year} {1999})}\BibitemShut {NoStop}%
\bibitem [{\citenamefont {Pendleton}\ and\ \citenamefont
  {Jensen}(2008)}]{pendleton2008compliant}%
  \BibitemOpen
  \bibfield  {author} {\bibinfo {author} {\bibfnamefont {T.~M.}\ \bibnamefont
  {Pendleton}}\ and\ \bibinfo {author} {\bibfnamefont {B.~D.}\ \bibnamefont
  {Jensen}},\ }\href@noop {} {\bibfield  {journal} {\bibinfo  {journal}
  {Journal of Mechanical Design}\ }\textbf {\bibinfo {volume} {130}},\ \bibinfo
  {pages} {122302} (\bibinfo {year} {2008})}\BibitemShut {NoStop}%
\bibitem [{\citenamefont {Cianchetti}\ \emph {et~al.}(2013)\citenamefont
  {Cianchetti}, \citenamefont {Ranzani}, \citenamefont {Gerboni}, \citenamefont
  {De~Falco}, \citenamefont {Laschi},\ and\ \citenamefont
  {Menciassi}}]{2013stiffflop}%
  \BibitemOpen
  \bibfield  {author} {\bibinfo {author} {\bibfnamefont {M.}~\bibnamefont
  {Cianchetti}}, \bibinfo {author} {\bibfnamefont {T.}~\bibnamefont {Ranzani}},
  \bibinfo {author} {\bibfnamefont {G.}~\bibnamefont {Gerboni}}, \bibinfo
  {author} {\bibfnamefont {I.}~\bibnamefont {De~Falco}}, \bibinfo {author}
  {\bibfnamefont {C.}~\bibnamefont {Laschi}}, \ and\ \bibinfo {author}
  {\bibfnamefont {A.}~\bibnamefont {Menciassi}},\ }in\ \href@noop {} {\emph
  {\bibinfo {booktitle} {Intelligent Robots and Systems (IROS), 2013 IEEE/RSJ
  International Conference on}}}\ (\bibinfo {organization} {IEEE},\ \bibinfo
  {year} {2013})\ pp.\ \bibinfo {pages} {3576--3581}\BibitemShut {NoStop}%
\bibitem [{\citenamefont {Filipov}\ \emph {et~al.}(2016)\citenamefont
  {Filipov}, \citenamefont {Paulino},\ and\ \citenamefont
  {Tachi}}]{filipov2016origami}%
  \BibitemOpen
  \bibfield  {author} {\bibinfo {author} {\bibfnamefont {E.~T.}\ \bibnamefont
  {Filipov}}, \bibinfo {author} {\bibfnamefont {G.}~\bibnamefont {Paulino}}, \
  and\ \bibinfo {author} {\bibfnamefont {T.}~\bibnamefont {Tachi}},\
  }\href@noop {} {\bibfield  {journal} {\bibinfo  {journal} {Proc. R. Soc. A}\
  }\textbf {\bibinfo {volume} {472}},\ \bibinfo {pages} {20150607} (\bibinfo
  {year} {2016})}\BibitemShut {NoStop}%
\bibitem [{\citenamefont {Kamrava}\ \emph {et~al.}(2018)\citenamefont
  {Kamrava}, \citenamefont {Mousanezhad}, \citenamefont {Felton},\ and\
  \citenamefont {Vaziri}}]{kamrava2018programmable}%
  \BibitemOpen
  \bibfield  {author} {\bibinfo {author} {\bibfnamefont {S.}~\bibnamefont
  {Kamrava}}, \bibinfo {author} {\bibfnamefont {D.}~\bibnamefont
  {Mousanezhad}}, \bibinfo {author} {\bibfnamefont {S.~M.}\ \bibnamefont
  {Felton}}, \ and\ \bibinfo {author} {\bibfnamefont {A.}~\bibnamefont
  {Vaziri}},\ }\href@noop {} {\bibfield  {journal} {\bibinfo  {journal}
  {Advanced Materials Technologies}\ }\textbf {\bibinfo {volume} {3}},\
  \bibinfo {pages} {1700276} (\bibinfo {year} {2018})}\BibitemShut {NoStop}%
\bibitem [{\citenamefont {Doedel}\ \emph {et~al.}(2007)\citenamefont {Doedel},
  \citenamefont {Paffenroth}, \citenamefont {Champneys}, \citenamefont
  {Fairgrieve}, \citenamefont {Kuznetsov}, \citenamefont {Oldeman},
  \citenamefont {Sandstede},\ and\ \citenamefont {Wang}}]{doedel2007auto}%
  \BibitemOpen
  \bibfield  {author} {\bibinfo {author} {\bibfnamefont {E.~J.}\ \bibnamefont
  {Doedel}}, \bibinfo {author} {\bibfnamefont {R.~C.}\ \bibnamefont
  {Paffenroth}}, \bibinfo {author} {\bibfnamefont {A.~R.}\ \bibnamefont
  {Champneys}}, \bibinfo {author} {\bibfnamefont {T.~F.}\ \bibnamefont
  {Fairgrieve}}, \bibinfo {author} {\bibfnamefont {Y.~A.}\ \bibnamefont
  {Kuznetsov}}, \bibinfo {author} {\bibfnamefont {B.~E.}\ \bibnamefont
  {Oldeman}}, \bibinfo {author} {\bibfnamefont {B.}~\bibnamefont {Sandstede}},
  \ and\ \bibinfo {author} {\bibfnamefont {X.}~\bibnamefont {Wang}},\
  }\href@noop {} {\enquote {\bibinfo {title} {{AUTO-07P}: Continuation and
  bifurcation software for ordinary differential equations},}\ }\bibinfo
  {howpublished} {\texttt{indy.cs.concordia.ca/auto/ }} (\bibinfo {year}
  {2007})\BibitemShut {NoStop}%
\bibitem [{\citenamefont {Cignoni}\ \emph {et~al.}(2008)\citenamefont
  {Cignoni}, \citenamefont {Callieri}, \citenamefont {Corsini}, \citenamefont
  {Dellepiane}, \citenamefont {Ganovelli},\ and\ \citenamefont
  {Ranzuglia}}]{meshlab_2008}%
  \BibitemOpen
  \bibfield  {author} {\bibinfo {author} {\bibfnamefont {P.}~\bibnamefont
  {Cignoni}}, \bibinfo {author} {\bibfnamefont {M.}~\bibnamefont {Callieri}},
  \bibinfo {author} {\bibfnamefont {M.}~\bibnamefont {Corsini}}, \bibinfo
  {author} {\bibfnamefont {M.}~\bibnamefont {Dellepiane}}, \bibinfo {author}
  {\bibfnamefont {F.}~\bibnamefont {Ganovelli}}, \ and\ \bibinfo {author}
  {\bibfnamefont {G.}~\bibnamefont {Ranzuglia}},\ }in\ \href@noop {} {\emph
  {\bibinfo {booktitle} {Eurographics Italian Chapter Conference}}},\ Vol.\
  \bibinfo {volume} {2008}\ (\bibinfo {year} {2008})\ pp.\ \bibinfo {pages}
  {129--136}\BibitemShut {NoStop}%
\bibitem [{\citenamefont {Corsini}\ \emph {et~al.}(2012)\citenamefont
  {Corsini}, \citenamefont {Cignoni},\ and\ \citenamefont
  {Scopigno}}]{meshlab_PC_2012}%
  \BibitemOpen
  \bibfield  {author} {\bibinfo {author} {\bibfnamefont {M.}~\bibnamefont
  {Corsini}}, \bibinfo {author} {\bibfnamefont {P.}~\bibnamefont {Cignoni}}, \
  and\ \bibinfo {author} {\bibfnamefont {R.}~\bibnamefont {Scopigno}},\
  }\href@noop {} {\bibfield  {journal} {\bibinfo  {journal} {IEEE Transactions
  on Visualization and Computer Graphics}\ }\textbf {\bibinfo {volume} {18}},\
  \bibinfo {pages} {914} (\bibinfo {year} {2012})}\BibitemShut {NoStop}%
\bibitem [{\citenamefont {Guennebaud}\ and\ \citenamefont
  {Gross}(2007)}]{apss_2007}%
  \BibitemOpen
  \bibfield  {author} {\bibinfo {author} {\bibfnamefont {G.}~\bibnamefont
  {Guennebaud}}\ and\ \bibinfo {author} {\bibfnamefont {M.}~\bibnamefont
  {Gross}},\ }in\ \href@noop {} {\emph {\bibinfo {booktitle} {ACM Transactions
  on Graphics (TOG)}}},\ Vol.~\bibinfo {volume} {26}\ (\bibinfo {organization}
  {ACM},\ \bibinfo {year} {2007})\ p.~\bibinfo {pages} {23}\BibitemShut
  {NoStop}%
\end{thebibliography}
\end{document}